\documentclass[12pt,a4paper]{article}
\counterwithin*{equation}{section}

\usepackage{caption}
\usepackage{cite,mathtools,amsmath,amssymb,graphicx,subfigure}
\usepackage[usenames]{color}
\usepackage[bookmarksopen,colorlinks=true,linkcolor=dark-green,citecolor=dark-red,urlcolor=dark-red,linktocpage=false]{hyperref}
\usepackage{url}
\usepackage{ulem}

\definecolor{dark-blue}{rgb}{0,0,0.6}
\definecolor{dark-green}{rgb}{0,0.4,0}
\definecolor{dark-red}{rgb}{0.6,0,0}

\usepackage[height=23.5cm,width=16.5cm,centering]{geometry}

\def\thefootnote{\fnsymbol{footnote}}
\newcommand{\gsim}{~\mbox{\raisebox{-1.0ex}{$\stackrel{\textstyle >}{\textstyle \sim}$ }}}
\newcommand{\lsim}{~\mbox{\raisebox{-1.0ex}{$\stackrel{\textstyle <}{\textstyle \sim}$ }}}

\newcommand{\beq}{\begin{align}}
\newcommand{\eeq}{\end{align}}
\newcommand{\beqa}{\begin{eqnarray}}
\newcommand{\eeqa}{\end{eqnarray}}
\newcommand{\mpl}{M_{\rm pl}}

\newcommand{\lmk}{\left(}
\newcommand{\rmk}{\right)}

\begin{document}
\begin{titlepage}

\begin{center}
	\hfill RESCEU-20/20 \\
	
	\vskip 10mm
	
	{\fontsize{20pt}{0pt} \bf
		Perturbative Reheating
	}
	\\[3mm]
	{\fontsize{20pt}{0pt} \bf
		in the Mixed Higgs-$ R^2 $ Model
	}
	
	\vskip 10mm
	
	{\large
		Minxi He\footnote{
			Email: {\tt he-minxi194"at"g.ecc.u-tokyo.ac.jp}
		}
	} 
	
	\vskip 5mm
	
	{\it Department of Physics, Graduate School of Science, The University of Tokyo,}
	\\[1mm]
	{\it Hongo 7-3-1, Bunkyo-ku, Tokyo 113-0033, Japan}
	\\[2mm]
	{\it Research Center for the Early Universe (RESCEU), Graduate School of Science,}
	\\[1mm]
	{\it The University of Tokyo, Hongo 7-3-1, Bunkyo-ku, Tokyo 113-0033, Japan}
	
\end{center}
\vskip 4mm

\begin{abstract}
The preheating process in the mixed Higgs-$ R^2 $ model has been investigated in depth recently, but the analysis of perturbative reheating is still missing. In this paper, we discuss the effect of perturbative decay during (p)reheating in this model. It is shown that perturbative decay can play an important role throughout the whole reheating process. Depending on the model parameters, perturbative decay can affect different stages of the reheating. We study the perturbative reheating with and without the presence of early preheating stage, and calculate the reheating temperature and the duration of the whole perturbative process. We find that the detail of the early preheating stage may not affect the final reheating temperature while it can affect the number of e-folds of reheating. 
	
\end{abstract}

\end{titlepage}

\tableofcontents
\thispagestyle{empty}
\newpage

\renewcommand{\thefootnote}{$\diamondsuit$\arabic{footnote}}
\setcounter{page}{1}
\setcounter{footnote}{0}

\section{Introduction}
\label{Sec-1}

For successful inflation models (see e.g. \cite{Sato:2015dga} for a review), reheating is essential  to connect the inflationary phase with the subsequent hot Big Bang cosmology. Reheating is a highly model-dependent process during which the energy that drives inflation is transferred to other matter fields by producing relativistic particles. Eventually, all the degrees of freedom in the Standard Model (SM) are populated and the universe is dominated by thermalized radiation. As an integral part of the cosmic history, the study of reheating can improve the observational constraints on inflation~\cite{Liddle:2003as,Martin:2010kz}\footnote{See Ref.~\cite{Gong:2015qha} for the Higgs inflation case.} by accurately expressing the pivot scale of curvature perturbation in terms of the $ e $-fold number of inflation. 

Reheating is generally characterized by two sequential stages for models where inflation is followed by inflaton oscillation. The first stage is preheating which is short but in some cases efficient to transfer energy from inflaton to relativistic particles through the dominant non-perturbative processes such as broad parametric resonance~\cite{Kofman:1994rk,Shtanov:1994ce,Kofman:1997yn,Greene:1997fu} or tachyonic instability~\cite{Felder:2000hj,Felder:2001kt,Kofman:2001rb} where the number of produced particles grows exponentially. The second stage is perturbative reheating during which the inflaton experiences perturbative decay and loses all the energy~\cite{Starobinsky:1980te,Abbott:1982hn,Dolgov:1982th}. 

This paper investigates the perturbative reheating in the mixed Higgs-$ R^2 $ model which involves a non-minimally coupled Higgs field and an $ R^2 $ term~\cite{Wang:2017fuy,Ema:2017rqn,He:2018gyf,Gundhi:2018wyz,Enckell:2018uic,Cheong:2019vzl}. This model is well-motivated in both phenomenological and theoretical aspects. It combines two observationally most favored inflationary models, the $R^2$-inflation~\cite{Starobinsky:1980te,Starobinsky:1983zz} and the original Higgs inflation with a non-minimal coupling between the Standard Model (SM) Higgs and gravity~\cite{CervantesCota:1995tz,Bezrukov:2007ep,Barvinsky:2008ia}\footnote{See Ref.~\cite{Kamada:2012se} for all the variants.}. In the case where isocurvature mode is heavy during inflation\footnote{The isocurvature mode of the attractor solution of inflation is heavy when $ \xi \gg 1 $ where $ \xi $ is the non-minimal coupling of Higgs defined in Eq.\eqref{eq-jordan-action} (see e.g. Ref.~\cite{He:2018gyf}). In this paper, we focus on this situation.}, this model inherits the predictions on the scalar spectral index $ n_s $ and tensor-to-scalar ratio $ r $ from its two single-field limits, the Higgs inflation and the $R^2$-inflation, which well match the observation by WMAP and Planck~\cite{Akrami:2018odb}. This is explained by the attractor behavior during inflation which enables us to describe this model as an effective $ R^2 $-inflation~\cite{He:2018gyf}. Depending on the model parameters, different reheating temperatures $ T_r $ can be realized which smoothly connects its two single-field limits. As the Higgs- and $R^2$-inflation can be distinguished by different reheating temperatures~\cite{Bezrukov:2011gp}, $ T_r $ is also applicable to break the degeneracy among different model parameters in the mixed Higgs-$ R^2 $ model. Conversely, more accurate observation in the future that determines $ T_r $ can settle down the mixing ratio between the Higgs- and $ R^2 $-inflation. 

From the theoretical point of view, the mixed Higgs-$ R^2 $ model can be a candidate of UV-extension of the Higgs inflation. The validity of pure Higgs inflation is questioned~\cite{Burgess:2009ea,Barbon:2009ya,Burgess:2010zq,Hertzberg:2010dc,Barvinsky:2009ii} due to the relatively low cutoff scale induced by the large non-minimal coupling, so is the prediction of this scenario~\cite{Bezrukov:2009db,Bezrukov:2014ipa}. Although it is argued that the unitarity is held during inflation~\cite{Bezrukov:2010jz}, the preheating stage is shown to enter the strongly coupled regime because of the large spiky effective mass of the Goldstone mode~\cite{DeCross:2015uza,JinnoThesis,Ema:2016dny,Sfakianakis:2018lzf}\footnote{See Ref.~\cite{Hamada:2020kuy} for the effect of higher dimensional operators on this phenomenon.}\footnote{The (p)reheating process in the Higgs inflation is also discussed in Refs.~\cite{Bezrukov:2008ut,GarciaBellido:2008ab,Repond:2016sol}.}. This issue can be resolved by the addition of an quadratic curvature term. It is pointed out by Refs.~\cite{Ema:2017rqn,Gorbunov:2018llf} that the cutoff scale of the Higgs inflation can be lifted up to Planck scale by the $ R^2 $ term. Furthermore, it has been found that the violent preheating mechanism in the pure Higgs inflation is largely weakened by the multi-field dynamics~\cite{He:2018mgb}. Analytical calculation of the spiky mass of the Goldstone mode and the energy density of produced particles shows that the spikes get much smaller and preheating is far less efficient than the single-field case. 

Therefore, the mixed Higgs-$ R^2 $ model is the most straightforward UV-extension of the Higgs inflation\footnote{Other proposal can be found in Refs.~\cite{Giudice:2010ka,Barbon:2015fla,Lee:2018esk}.}. The presence of the $ R^2 $ term is a natural consequence in the renormalization group running~\cite{Salvio:2015kka,Netto:2015cba,Calmet:2016fsr,Liu:2018hno,Ghilencea:2018rqg}. Additionally, the origin of the quadratic curvature term is discussed in Refs.~\cite{Ema:2019fdd,Ema:2020zvg} in terms of scattering amplitude and the non-linear sigma model. 

At the reheating epoch in the mixed Higgs-$ R^2 $ model, the effective single-field description breaks down and the multi-field dynamics of the Higgs $ h $ and scalaron $ \varphi $ becomes crucial to determine the evolution of the system. The preheating process of the mixed Higgs-$ R^2 $ model has been investigated extensively~\cite{He:2018mgb,Bezrukov:2019ylq,He:2020ivk,Bezrukov:2020txg}\footnote{Preheating process in similar models is studied in, e.g. Ref.~\cite{Fu:2019qqe}.}. As mentioned above, Ref.~\cite{He:2018mgb} showed that the first stage of preheating is inefficient to have significant effect on the whole reheating process due to the presence of the $ R^2 $ term. Later, Ref.~\cite{Bezrukov:2019ylq} found that the physical Higgs and the longitudinal mode of weak gauge bosons can experience tachyonic instability right after the first stage with critical choice of model parameters in the Higgs-like regime (see Sec.~\ref{Sec-2} for the definition). Actually, there is a bifurcation point at the origin of the two dimensional potential that divides the $ \varphi >0 $ regime into three parts, a potential hill at $ h=0 $ and two valleys where inflation occurs as an attractor (see Fig.~\ref{fig-potential}). The tachyonic instability is induced by this potential hill if the inflaton oscillation after the end of inflation enters such a regime. In that case, the inhomogeneity of Higgs and the longitudinal mode of weak gauge bosons grow exponentially, namely the occurrence of tachyonic preheating\footnote{Tachyonic preheating was first studied in theories with spontaneous symmetry breaking~\cite{Felder:2000hj,Felder:2001kt,Kofman:2001rb}. Tachyonic instability for the spectator field can also be induced by large field space curvature, see, e.g. Ref.~\cite{Iarygina:2020dwe}.}. Consequently, the dynamics of the whole system during reheating becomes chaotic in the sense described in Ref.~\cite{Podolsky:2002qv,Jin:2004bf}. In order to quantitatively understand the distribution of the critical parameters that induce tachyonic preheating, the efficiency of such instability, and how much fine-tuning is needed, analytical calculation is presented in Ref.~\cite{He:2020ivk} showing that the critical parameters appears in both Higgs- and $ R^2 $-like regimes. The phase of the Higgs oscillation during $ \varphi <0 $ determines the positions of the critical parameters that can realize locally maximal strength in the phase space. With the method proposed in Ref.~\cite{He:2018mgb}, the analytical calculation of tachyonic particle production was carried out and estimation of the degree of fine-tuning needed to complete preheating solely by tachyonic preheating was also obtained by combining the analytical and numerical methods. The results in Ref.~\cite{He:2020ivk} are consistent with the elaborate numerical calculation in Ref.~\cite{Bezrukov:2019ylq,Bezrukov:2020txg} and further allow us to determine the efficiency depending on the tachyonic duration for a given set of model parameters in the whole parameter space. 

However, the backreaction from the produced particles was not taken into account in Ref.~\cite{He:2020ivk}. One would expect that the backreaction is possible to terminate the tachyonic instability midway so that subsequent non-tachyonic mechanism should be considered to continue the reheating of the universe. In the lattice simulation carried out in Ref.~\cite{Bezrukov:2020txg}, the tachyonic preheating is studied by taking into account the backreaction from the produced particles as a classical rescattering effect between the homogeneous background and the inhomogeneous modes in the $ R^2 $-like regime. The strength of the rescattering also depends on the model parameter, getting stronger for more Higgs-like parameters and weaker for more $ R^2 $-like choices. After a few hundreds scalaron oscillations, it was found that some fraction of the homogeneous background can still remain apart from the relativistic particles. Since radiation is redshifted much faster than the homogeneous background field by cosmic expansion, one would expect that the homogeneous part can once again dominate the universe and finally decay away perturbatively. The observation indicates a possibility that the reheating temperature $ T_r $ may be only determined by the final perturbative process regardless of the details of preheating, which motivates us to further investigate the perturbative reheating in this model. 

The structure of this paper is as follows. In the next section, we briefly review the mixed Higgs-$ R^2 $ model and investigate the background evolution of Higgs and scalaron analytically, which will be helpful in the rest parts of the paper. In Sec.~\ref{Sec-3}, we discuss the role played by perturbative reheating in the whole reheating process with and without tachyonic preheating. We will also calculate the reheating temperature for different model parameters. Note that we will not directly calculate tachyonic effect in this paper. We give a summary in Sec.~\ref{Sec-4} where we also discuss the caveats of this work and the future outlook. 

\section{Mixed Higgs-\texorpdfstring{$ R^2 $}{Lg} model and background analysis}
\label{Sec-2}

\subsection{Brief review of the mixed Higgs-\texorpdfstring{$ R^2 $}{Lg} model}\label{Subsec-model}

The Lagrangian of the mixed Higgs-$R^2$ model is originally given in the Jordan frame~\cite{Wang:2017fuy,Ema:2017rqn,He:2018gyf,Gundhi:2018wyz}
\begin{align}\label{eq-jordan-action}
    S_{\mathrm J}
    &= \int\!\! d^4 x \sqrt{-g_{\mathrm J}} {\cal L}_\mathrm{J} \notag \\
    &=
    \int\!\! d^4 x \sqrt{-g_{\mathrm J}} 
    \left[ \left(\frac{M_{\mathrm{pl}}^2}{2} + \xi |{\mathcal H}|^2 \right)R_{\mathrm J} + \frac{M_{\mathrm{pl}}^2}{12 M^2}R_{\mathrm J}^2 
    - g_{\mathrm J}^{\mu\nu} \partial_\mu  {\mathcal H} \partial_\nu {\mathcal H}^\dagger - \lambda |{\mathcal H}|^4 \right] ~,  
\end{align}
with the Ricci scalar $ R_{\mathrm J} $, reduced Planck mass $M_\mathrm{pl}$, and the SM Higgs $ {\mathcal H} $ where the subscript~``J" is used to identify the quantities defined in Jordan frame. The mass term of Higgs is neglected because the energy scale of inflation and reheating is much higher than the electroweak scale. Therefore, there are three model parameters, the scalaron mass $ M $, the non-minimal coupling $ \xi>0 $\footnote{The conformal coupling corresponds to $ \xi =-1/6 $. The $ \xi <0 $ case is studied e.g. in \cite{Pi:2017gih}.}, and the self-coupling of Higgs $ \lambda $. Throughout this paper, we only consider stable and non-critical SM Higgs up to inflationary scale so we will take $ \lambda =0.01 $ as a constant~\cite{Bezrukov:2009db,Bezrukov:2014ipa}.

By a conformal transformation \cite{Maeda:1987xf,Maeda:1988ab}
\begin{align}\label{eq-conformal-transformation}
    g_{\mathrm{E}\mu \nu} (x)=e^{\sqrt{\frac{2}{3}}\frac{\varphi(x)}{\mpl}} {g_\mathrm{J}}_{\mu \nu}(x)
    \equiv e^{\alpha\varphi(x)}{g_\mathrm{J}}_{\mu \nu}(x) ~,
\end{align}
where we define the ``scalaron'' field $\varphi$ as
\begin{align}\label{def:psi}
    \sqrt{\frac{2}{3}}\frac{\varphi}{\mpl}\equiv \ln \lmk\frac{2}{\mpl^2}\left|\frac{\partial {\cal L}_\mathrm{J}}{\partial {R_\mathrm{J}}}\right|\rmk ~,
\end{align}
the action \eqref{eq-jordan-action} can be transformed into Einstein frame written as 
\begin{align}
    \label{eq-einstein-action}
	S_{\mathrm{E}} 
	&=
	\int d^4 x \sqrt{-g_{\mathrm E}} 
	\left[\frac{M_{\mathrm{pl}}^2}{2} R_{\mathrm{E}} -\frac{1}{2} g_{\mathrm E}^{\mu\nu}\partial_\mu \varphi \partial_\nu \varphi 
	- e^{-\alpha \varphi} g^{\mu\nu}_{\mathrm E} \partial_\mu {\mathcal H}^{\dagger} \partial_\nu {\mathcal H} -U(\varphi, {\mathcal H}) \right] ~,
	\\ 
	\label{eq-einstein-potential}
	U(\varphi, {\mathcal H}) 
	&= 
	\lambda e^{-2 \alpha \varphi} |{\mathcal H}|^4  + \frac{3}{4} M_{\mathrm{pl}}^2 M^2 
	\left[1 - \left(1 + \frac{2\xi}{M_{\mathrm{pl}}^2} |{\mathcal H}|^2\right)e^{-\alpha \varphi} \right]^2 
\end{align}
where the subscript~``E" is used to identify the quantities defined in Einstein frame but will be omitted hereafter for convenience because we will work in Einstein frame from now on. The physics is not changed by the transformation~\eqref{eq-conformal-transformation}. 

It is shown in Ref.~\cite{He:2018gyf} that the inflation dynamics of this model can be described by effective Starobinsky model with a modified scalaron mass 
\begin{align}\label{eq-observation-constraint}
    \tilde{M}^2 \equiv \frac{M^2}{1+ \frac{3\xi^2}{\lambda} \frac{M^2}{M^2_{\rm pl}}} =M^2_c
\end{align}
where $ M_c \simeq 1.3 \times 10^{-5} M_{\rm pl} $ which is fixed by the amplitude of temperature fluctuations of cosmic microwave background (CMB). We will use $ \tilde{M} $ and $ M_c $ interchangeably in this paper. This relation reduces the number of parameters so that one can use either $ M $ or $ \xi $ to characterize the system. It is also useful to rewrite the relation above as 
\begin{align}
    \frac{\xi^2}{\xi^2_c} + \frac{M^2_c}{M^2} =1
\end{align}
where the other critical value is given by $ \xi_c = \sqrt{\lambda M^2_{\rm pl}/3M^2_c} \simeq 4.4\times 10^3 $. For later convenience, one can define a new parameter $ 0\leq \theta \leq \pi/2 $ such that 
\begin{align}\label{eq-def-theta}
	\cos \theta = \frac{\xi}{\xi_c} ~,~~~\sin \theta= \frac{M_c}{M}
\end{align}
where the Higgs-like regime is covered by $ 0 < \theta < \pi/4 $ while the $ R^2 $-like regime by $ \pi/4 < \theta < \pi/2  $. 

The parameter space of the mixed Higgs-$ R^2 $ model is separated into three parts, strong coupling, Higgs-like, and $ R^2 $-like regimes. We keep our discussion outside the strong coupling regime, namely, in the region 
\begin{equation} \label{eq-weakm}
    M \lsim \sqrt{\frac{4 \pi}{3}} \frac{M_\mathrm{pl}}{\xi}, 
\end{equation}
so that the system is weakly coupled below $ M_{\rm pl} $, which is also an advantage of this model as a UV-extension of the Higgs inflation because the strong coupling scale in the pure Higgs inflation is much lower $ \Lambda \simeq M_\mathrm{pl}/\xi $. Equations~\eqref{eq-observation-constraint} and \eqref{eq-weakm} determines the boundary $ \xi_s  \equiv 1/\sqrt{1/\xi_c^2+3M_c^2/4 \pi M_\mathrm{pl}^2} \cong \xi_c (1-\lambda/8\pi) \simeq 4.4 \times 10^{3} $ which is slightly smaller than $ \xi_c $. The Higgs-like regime is within $ \xi_c/\sqrt{2} \simeq  3.1\times 10^3 \lsim \xi \lsim   \xi_s $, where the former inequality comes from the condition $\xi/\xi_c > M_c/M$. The $ R^2 $-like regime is then for $ 0 \leq \xi \lsim  \xi_c/\sqrt{2} $. 

\subsection{Background evolution}\label{Subsec-bg}

Since we focus on the perturbative reheating in this paper, only the non-critical parameter choice is considered here. To be specific, ``non-critical parameter" means the parameters cannot realize tachyonic instability at the first scalaron oscillation right after the end of inflation. The detailed analytical study of the background evolution in the critical case where tachyonic preheating is realized at the first scalaron oscillation is given in Ref.~\cite{He:2020ivk}. 

\begin{figure}[h]
	\centering
	\includegraphics[width=.5\textwidth]{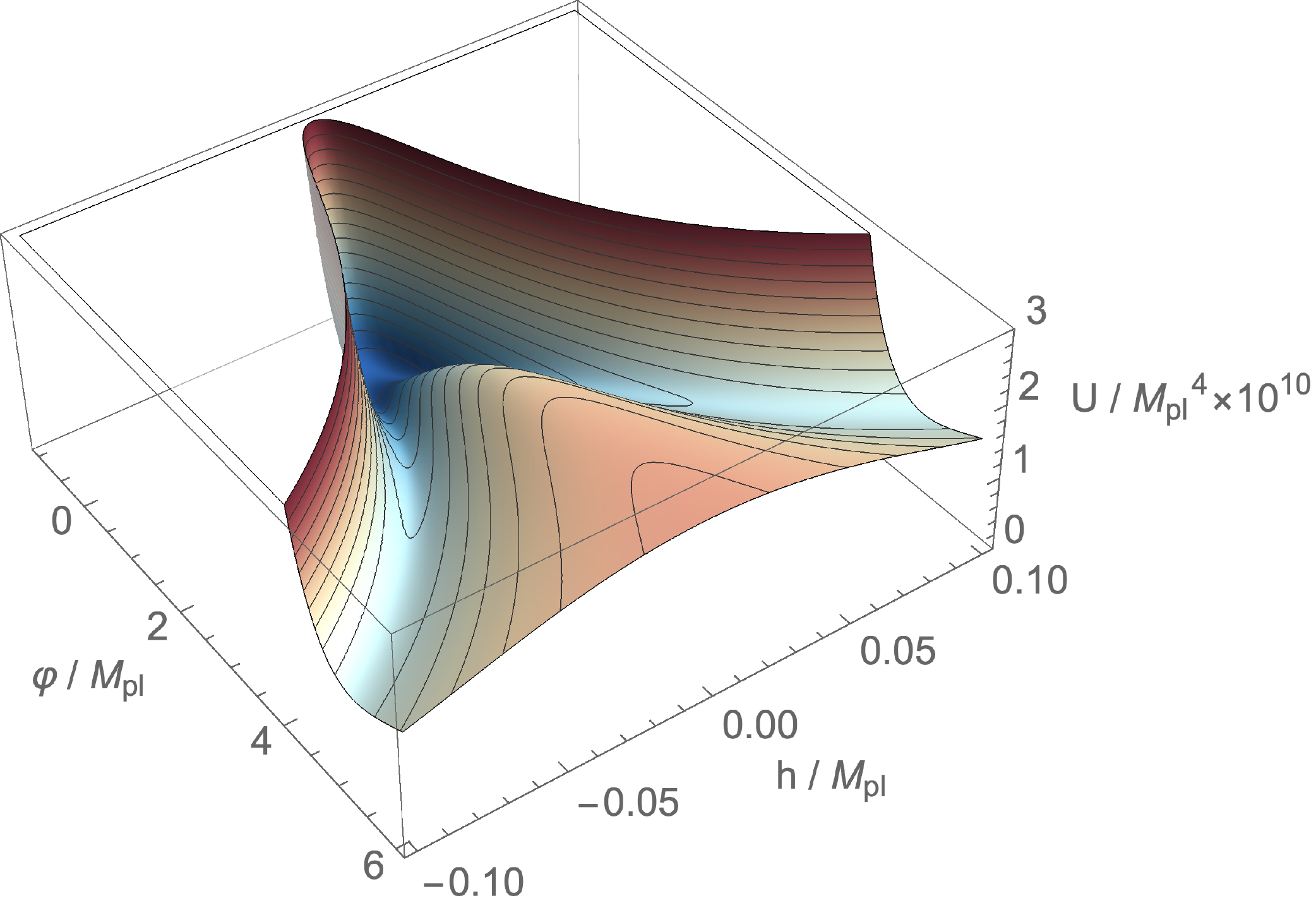}
	\caption{An example for the shape of the potential. The parameters are chosen as $\lambda = 0.01$ and  $ \xi=3000 $ with Eq.~\eqref{eq-observation-constraint} satisfied.}
	\label{fig-potential}
\end{figure}

The typical potential shape in this model is shown in Fig.~\ref{fig-potential}. 
As can be seen there, there is only one local minimum in $ \varphi<0 $ regime, while there are two valleys in $ \varphi >0 $ regime where the attractor behavior during inflation takes place, sandwiching a potential hill induced by the non-minimal coupling $ \xi $. After the end of inflation, scalaron oscillates around the origin where the bifurcation point provides basically three paths for the Higgs field upon the entry to $ \varphi >0 $ regime, entering one of the valleys or staying around $ h=0 $. The latter corresponds to the critical case where tachyonic preheating occurs~\cite{Bezrukov:2019ylq,He:2020ivk,Bezrukov:2020txg}. In the former case, for parameters close to the critical choices, large amplitude of Higgs oscillation around the valley is expected, which implies that large amount of energy is transferred from scalaron to Higgs. On the contrary, if the parameter choice is far from the critical case, the amplitude of coherent Higgs oscillation is small so there is little energy stored in Higgs background. See Appendix~\ref{Appendix-kinetic-Higgs} for more discussion. Later, this observation will play an important role when we determine the reheating temperature by perturbative decay. 

\begin{figure}[h]
	\centering
	\begin{subfigure}
		\centering
		\includegraphics[width=.48\textwidth]{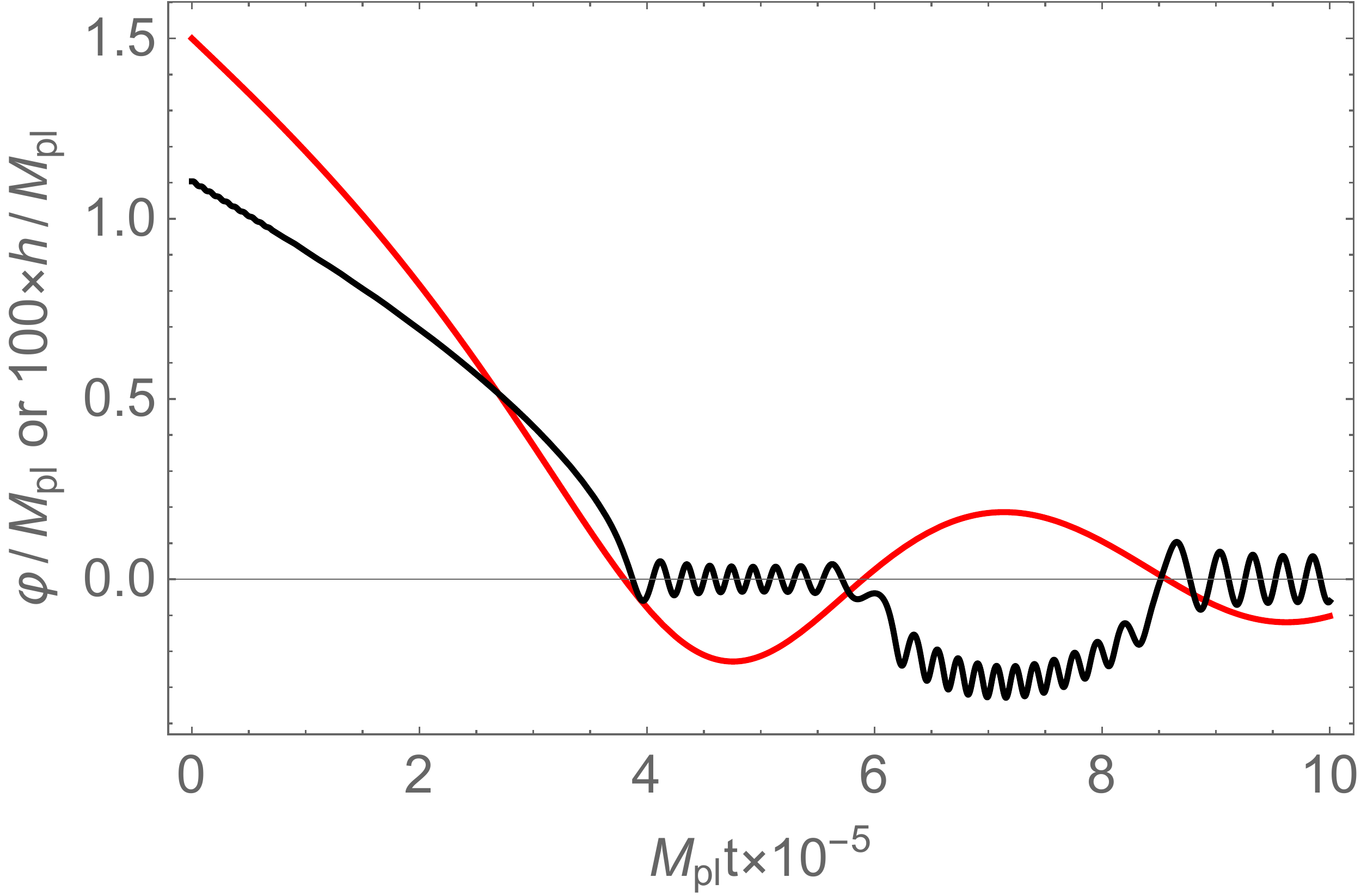}
	\end{subfigure}
	\begin{subfigure}
		\centering
		\includegraphics[width=.48\textwidth]{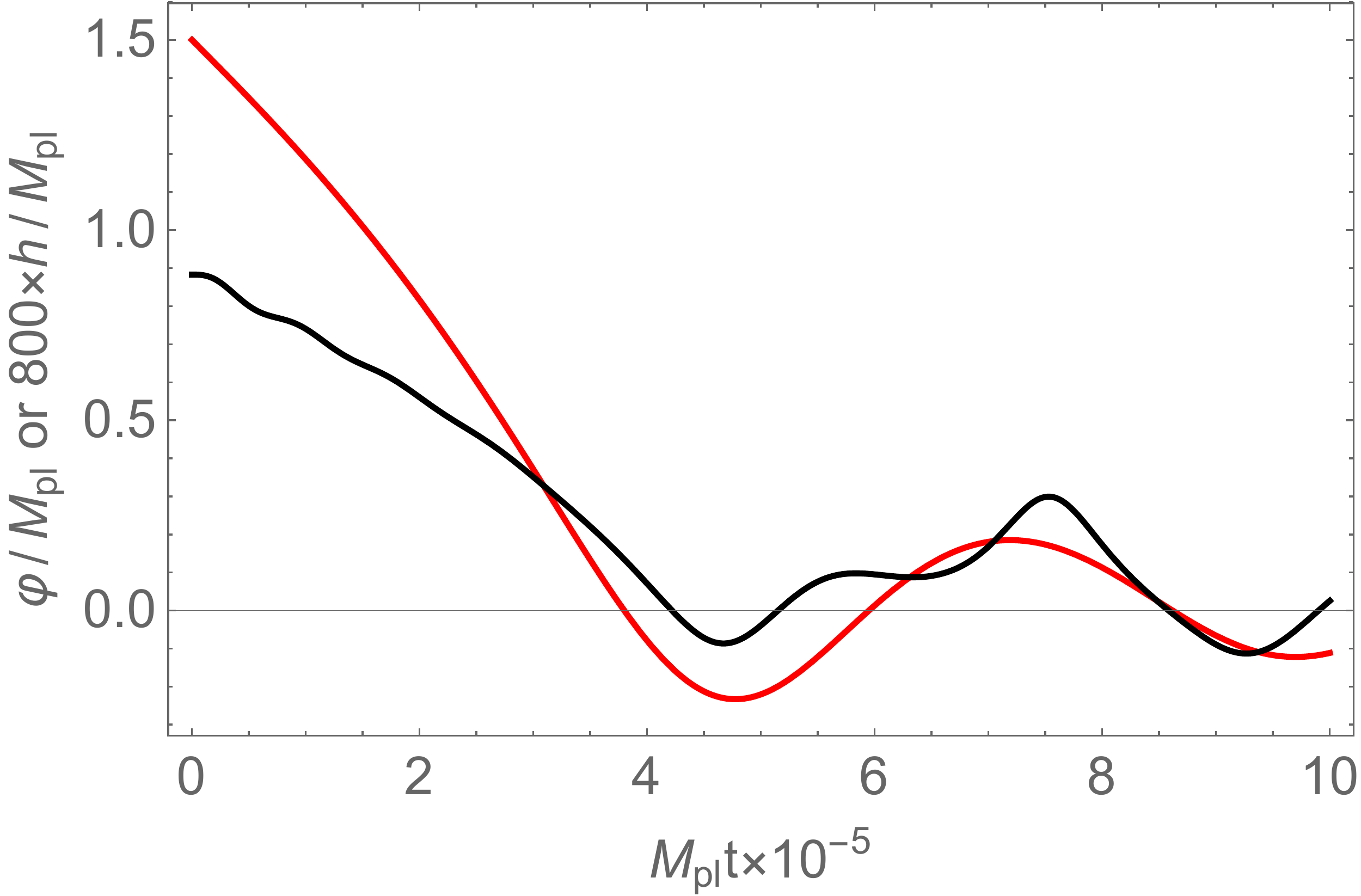}
	\end{subfigure}
	\caption{Background dynamics of the scalaron (red) and Higgs (black). Left: $ \xi=1000 $. Right: $ \xi=10 $. When $ \xi \lsim \mathcal{O}(10) $, the frequency of the small oscillation of Higgs is comparable with the scalaron frequency so the small oscillations ``disappear" on the right panel.}
	\label{fig-scalaron-higgs-bg}
\end{figure}

Next, we are going to analyze the background evolution of Higgs and scalaron, respectively. Note that we work in the unitary gauge where $ \mathcal{H} $ in the Lagrangian can be simplified to be a real scalar $ h/\sqrt{2} $ representing the one physical degree of freedom. This is sufficient for background-level analysis. For the convenience of the following discussion, we show two examples of the behavior of the homogeneous scalaron and Higgs in Fig.~\ref{fig-scalaron-higgs-bg}. 
It can be easily seen that the hierarchy of oscillation frequencies between Higgs and scalaron is large for $ \xi \gg \mathcal{O}(10) $ while small for $ \xi \lsim \mathcal{O}(10) $. In the former case, WKB approximation is valid because scalaron is approximately constant within one Higgs oscillation. On the contrary, this approximation breaks down for the latter so analytical description is difficult to find. Therefore, we will only give analysis for $ \xi \gg \mathcal{O}(10) $. For $ \xi \lsim \mathcal{O}(10) $, we expect that the situation quick approaches to that of $ R^2 $-inflation as $ \xi $ decreases but we will not discuss it in detail. Also note that we will not consider critical cases which have already been analytically investigated in Ref.~\cite{He:2020ivk}. 

\subsubsection{Homogeneous Higgs field}\label{Subsubsec-bg-Higgs}

During the reheating epoch, the homogeneous $ h $ approximately traces the local minima (or the valleys) in the Higgs direction of the potential 
\begin{align}\label{eq-h-valley}
    h^2_{\rm v} = \left\{
    \begin{array}{ll}
        0 &~, \varphi <0 \\ [0.3cm]
        \frac{3\xi (e^{\alpha \varphi} -1)}{\lambda} \tilde{M}^2 \approx \frac{3\xi \alpha}{\lambda} M^2_c \varphi &~, \varphi>0 
    \end{array}
    \right.
\end{align}
with small oscillation around it. If the amplitude of this oscillation is too large, the effective mass of the Higgs field might be tachyonic, which is beyond the scope of this paper. This constraint simplifies the situation when studying the non-critical case where Higgs oscillates along the valleys in $ \varphi >0 $ regime. Apart from Eq.~\eqref{eq-h-valley}, the small oscillations around the valleys are essential. We can separate the Higgs into two parts as 
\begin{align}\label{eq-higgs-valley-plus-oscillation}
    h(t) \approx h_{\rm v}(t) +h_{\rm osc} (t) 
\end{align}
where $ h_{\rm osc} $ represents the small oscillation part of homogeneous Higgs. This separation is not well-defined around the transition moments when $ \varphi $ crosses zero but our interest is not on these regimes. 

The background equation of motion for Higgs is 
\begin{align}\label{eq-bgeom-higgs-1}
    \ddot{h} +3H \dot{h} +e^{\alpha \varphi} \frac{\partial U}{\partial h} -\alpha \dot{\varphi} \dot{h} =0 ~.
\end{align}
Note that for most part of the parameter space $ \xi \gg \mathcal{O}(10) $, $ |\dot{h}| \gg |\dot{\varphi}| $ as mentioned previously. Under such condition, the time-dependence of $ h_{\rm osc} $ are much more significant than those of $ h_{\rm v} $ for all time. In other words, one can approximate $ \dot{h} \approx \dot{h}_{\rm osc} $ and $ \ddot{h} \approx \ddot{h}_{\rm osc} $. In addition, $ |h_{\rm v}| $ dominates over $ |h_{\rm osc}| $ most of the time during $ \varphi >0 $ in the situation we consider here except for, again, around the transition moments\footnote{Similar situation is considered in Ref.~\cite{Fan:2019udt}.}. Otherwise, $ h $ can switch between positive and negative values during the oscillation or enter the tachyonic regime as mentioned above, which will not be discussed in this paper. 

In the case with $ \xi \gg \mathcal{O}(10) $, we consider the third term in Eq.~\eqref{eq-bgeom-higgs-1}, keeping Eq.~\eqref{eq-h-valley} and $ |h_{\rm v}| \gg |h_{\rm osc}| $ when $ \varphi >0 $ in mind, 
\begin{align}
    e^{\alpha \varphi} \frac{\partial U}{\partial h} \left\{ 
    \begin{array}{ll}
        =\lambda \left( 1+ \frac{3\xi^2 M^2}{\lambda M^2_{\rm pl}} \right) e^{-\alpha \varphi} h^3_{\rm osc} -3\xi M^2\left( 1-e^{-\alpha \varphi} \right) h_{\rm osc} \approx -3\xi M^2 \alpha \varphi h_{\rm osc} &, \varphi <0 \\ [0.3cm]
        \approx 6\xi M^2 \alpha \varphi h_{\rm osc} &, \varphi >0 ~.
    \end{array}
    \right.
\end{align}
As a result, one obtains 
    \begin{align}\label{eq-bgeom-higgs-2}
    \ddot{h}_{\rm osc} +3H \dot{h}_{\rm osc} + m^2_{h_{\rm osc}} h_{\rm osc} \approx 0
\end{align}
where the last term in Eq.~\eqref{eq-bgeom-higgs-1} is neglected (because it is Planck-suppressed and $ \dot{\varphi} $ is small) and 
\begin{align}\label{eq-higgs-effmass}
    m^2_{h_{\rm osc}} \equiv \left\{
    \begin{array}{ll}
        -3\xi M^2 \alpha \varphi &, \varphi <0 \\ [0.3cm]
        6\xi M^2 \alpha \varphi &, \varphi >0 ~.
    \end{array}
    \right.
\end{align} 
Obviously, the effective mass of $ h_{\rm osc} $ depends on $ \varphi (t) $. For $ \xi \gg \mathcal{O}(10) $,  $ m^2_{h_{\rm osc}} \gg m^2_{\varphi} $ so that Higgs oscillates many times within one scalaron period, as in the left panel of Fig.~\ref{fig-scalaron-higgs-bg}. As a result, with WKB approximation, one can assume the form of $ h_{\rm osc} $ as (see Appendix~\ref{Appendix-analytical-bg} for detailed explanation)
\begin{align}\label{eq-bg-higgs-WKB}
    h_{\rm osc} = \frac{A_h}{\sqrt{2m_{h_{\rm osc}}}} \cos\left[\int^t m_{h_{\rm osc}} (t') dt' +\delta_i \right] 
\end{align}
where we need the evolution of $ \varphi $ to finish the calculation. The phase $ \delta_i $ and $ A_h $ are determined by the initial condition. The detail discussion on $ \varphi $ is done in Sec.~\ref{Subsubsec-bg-scalaron} but now we make use of the conclusion there to derive $ h_{\rm osc} $. Neglecting the cosmic expansion, $ \varphi \propto \sin (Mt) $ or $ \varphi \propto \sin (\tilde{M}t) $. As a result, 
\begin{align}\label{eq-bg-higgs-analytic}
    h_{\rm osc} = \left\{
    \begin{array}{ll}
        \frac{A_h}{\sqrt{2m_{h_{\rm osc}}}} \cos\left( -2 \sqrt{3\xi \alpha |\varphi_a|} E\left[ \frac{\pi}{4} -\frac{Mt}{2},2 \right] +\delta_i \right) &, \varphi <0 \\ [0.3cm]
        \frac{A_h}{\sqrt{2m_{h_{\rm osc}}}} \cos\left( -2 \sqrt{6\xi \alpha \varphi_a} \frac{M}{\tilde{M}} E\left[ \frac{\pi}{4} -\frac{\tilde{M}t}{2}, 2 \right] +\delta_i \right) &, \varphi >0 
    \end{array}
    \right. 
\end{align}
which is shown to be a good approximation in Appendix~\ref{Appendix-analytical-bg} where we have defined $ \varphi_a $ to be the amplitude of scalaron oscillation $ \varphi =\varphi_a \sin \left( M(\varphi) t \right) $ for both positive and negative $ \varphi $. 

\subsubsection{Homogeneous scalaron field}\label{Subsubsec-bg-scalaron}

Putting Eq.~\eqref{eq-h-valley} back to Eq.~\eqref{eq-einstein-potential},  we have an effective potential for the homogeneous scalaron 
\begin{align}
    U_{\rm v} (\varphi) =\left\{
    \begin{array}{ll}
        \frac{3}{4} M^2_{\rm pl} M^2 (1-e^{-\alpha \varphi})^2 \approx \frac{1}{2} M^2 \varphi^2 &, \varphi<0 \\ [0.3cm]
        \frac{3}{4} M^2_{\rm pl} \tilde{M}^2 (1-e^{-\alpha \varphi})^2 \approx \frac{1}{2} \tilde{M}^2 \varphi^2 &, \varphi >0
    \end{array}
    \right. 
\end{align}
where the approximations are for small-field expansion around $ \varphi =0 $ that is valid during reheating. As can be seen, the scalaron simply oscillates harmonically but with different masses for $ \varphi>0 $ and $ \varphi <0 $. The background equation for $ \varphi $ is given by 
\begin{align}\label{eq-bgeom-scalaron-1}
    \ddot{\varphi} +3H \dot{\varphi} + \frac{\partial U}{\partial \varphi} +\frac{\alpha}{2} e^{-\alpha \varphi} \dot{h}^2 =0
\end{align}
where the third term can be simplified by Eq.~\eqref{eq-h-valley} as 
\begin{align}
    \frac{\partial U}{\partial \varphi} =\left\{
    \begin{array}{ll}
        \frac{M^2}{\alpha} \left( 1- e^{-\alpha \varphi} \right) e^{-\alpha \varphi} \approx M^2\varphi &, \varphi <0 \\[0.3cm]
        \frac{\tilde{M}^2}{\alpha} \left( 1- e^{-\alpha \varphi} \right) e^{-\alpha \varphi} \approx \tilde{M}^2\varphi &, \varphi >0 ~.
    \end{array}
    \right.
\end{align}
As a result, the scalaron evolves as 
\begin{align}\label{eq-bgeom-scalaron-2}
    \ddot{\varphi} +3H \dot{\varphi} + m^2_{\varphi} \varphi \approx 0
\end{align}
where we have neglected the Planck-suppressed term and the time-dependent mass is given by 
\begin{align}\label{eq-scalaron-effmass}
    m^2_{\varphi} \equiv \left\{
    \begin{array}{ll}
        M^2 &, \varphi<0 \\[0.3cm]
        \tilde{M}^2 &, \varphi>0 ~.
    \end{array}
    \right.
\end{align}
Note that when $ \xi \lsim 1800 $, $ (M-\tilde{M})/\tilde{M} \lsim 10 \% $ ~so $ \varphi $ is basically oscillating with an unchanged frequency. Only when we are in the Higgs-like regime, the difference between $ M $ and $ \tilde{M} $ is significant. Since the universe is matter-dominated before the completion of reheating due to the fact that the oscillating scalar field acts as non-relativistic matter, the Hubble parameter is simply $ H \approx \frac{2}{3t} $. As a result, Eq.~\eqref{eq-bgeom-scalaron-2} can be solved analytically for either $ \varphi >0 $ or $ \varphi<0 $
\begin{align}\label{eq-bgeom-scalaron-separate-solution}
    \varphi (t) =\left\{
    \begin{array}{ll}
        \varphi_1 \frac{\sin(Mt)}{Mt} &, \varphi<0 \\ [0.3cm]
        \varphi_2 \frac{\sin(\tilde{M}t)}{\tilde{M}t} &, \varphi >0
    \end{array}
    \right.
\end{align}
where we have required $ \varphi $ to be regular at $ t=0 $. $ \varphi_1<0 $ and $ \varphi_2>0 $ are the amplitudes for each period by the initial condition and they can be determined by the same method used in Refs.~\cite{He:2018mgb,He:2020ivk}. 

So much is the discussion of background evolution without incorporating any dissipative effects on the Higgs and scalaron fields. 

\section{Role of perturbative reheating in mixed Higgs-$ R^2 $ model}
\label{Sec-3}

In this section, we will calculate the decay rate of both background fields, scalaron and Higgs. As will be seen later, the dominant decay channel for scalaron is to decay into Higgs particles while the Higgs field mainly decays into the heavy particles in SM, such top and bottom quarks, $ W^{\pm} $ and $ Z $ bosons, depending on the model parameters. The decay rate of the inhomogeneous Higgs will also be estimated because it can play an important role when discussing their backreaction on the homogeneous background. Based on these results, we will show how the perturbative decay affects the whole reheating process in the mixed Higgs-$ R^2 $ model, including the cases with and without tachyonic preheating. Again, we do not directly analyze tachyonic particle production or relevant effects. 

\subsection{Decay rate}\label{Subsec-decay-rate}

In order to estimate the perturbative decay rate of scalaron and Higgs, it is necessary to write down the relevant interaction terms among them and SM particles. In the Jordan frame, we assume that all SM fields except for Higgs are minimally coupled with gravity. After the conformal transformation to the Einstein frame, the scalaron couples with these matter fields through the conformal factor which is Planck-suppressed. In the small $ \varphi $ limit, it can be neglected at the leading order. Therefore, the dominant channel for scalaron should be the decay into Higgs particles through the non-minimal coupling term. Of course, it is well-known that such decay exists even when $ \xi =0 $, although it is much less efficient than large $ \xi $ case. On the other hand, the Higgs field directly couples not only with scalaron through the non-minimal coupling but other SM particles such as quarks and weak bosons with large coupling constants, which allows Higgs to decay much more efficiently. In this paper, we do not take into account the interactions between gauge sector and fermion sector because the main goal is to deplete the energy of the homogeneous background. 

As mentioned above, for inflation it is sufficient to use the unitary gauge. During reheating, however, the homogeneous Higgs field oscillates around the origin as can be seen in Eq.~\eqref{eq-h-valley}. It is known that the unitary gauge is not well defined when $ h=0 $ so one may need to work in other well-defined gauges when studying reheating process. However, as shown in Ref.~\cite{Ema:2016dny}, crossing $ h=0 $ does not spoil the calculation in the unitary gauge. Therefore, we will continue using the unitary gauge in the following calculation. The derivation of the decay rates for both Higgs and scalaron is presented in Appendix~\ref{Appendix-decay-rate}. We will directly use the results here. 

Before proceeding, we clarify the notations used below. As in Appendix~\ref{Appendix-decay-rate}, we separate the Higgs field into homogeneous and inhomogeneous parts $ h=h_0(t) +\delta h(t,x) $ where $ h_0(t) $ is given by Eq.~\eqref{eq-higgs-valley-plus-oscillation}. And we redefine the fields $ \tilde{\delta h} \equiv a^{3/2} \delta h $, $ \tilde{W}^{\pm}_{\mu} \equiv a^{3/2} W^{\pm}_{\mu} $, $ \tilde{Z}_{\mu} \equiv a^{3/2} Z_{\mu} $, $ \tilde{t} \equiv a^{3/2} t $, and $ \tilde{b} \equiv a^{3/2} b $ in the decay rate calculation. 

The decay of Higgs considered here is mainly through two channels, to weak gauge bosons and to top and bottom quarks due to the large coupling constants. The gauge boson channels are only opened in the $ \varphi >0 $ regime when the conditions 
\begin{align}
    \xi^2 >& \xi_c^2 \left( 1-\frac{2\lambda}{g^2} \right) \equiv \xi_W \simeq 4291 \\
    \xi^2 >& \xi_c^2 \left( 1-\frac{2\lambda}{g^2+g'^2} \right) \equiv \xi_Z \simeq 4347 
\end{align}
are satisfied for $ W^{\pm} $ and $ Z $ respectively, which means that it is allowed only in the deep Higgs-like regime. The decay rates are given by 
\begin{align}
	\Gamma_{h_0 \rightarrow \tilde{W}\tilde{W}} &= \frac{\lambda}{8\pi} \frac{M^3}{\tilde{M}^2} \left( 6\xi \alpha\varphi \right)^{1/2} \left( 1-\frac{g^2}{2\lambda} \frac{\tilde{M}^2}{M^2} + \frac{3g^4}{16\lambda^2} \frac{\tilde{M}^4}{M^4} \right) \left( 1-\frac{g^2}{2\lambda} \frac{\tilde{M}^2}{M^2} \right)^{1/2} \\
	\Gamma_{h_0 \rightarrow \tilde{Z}\tilde{Z}} &= \frac{\lambda}{16\pi} \frac{M^3}{\tilde{M}^2} \left( 6\xi \alpha\varphi \right)^{1/2} \left( 1- \frac{g^2+g'^2}{2\lambda} \frac{\tilde{M}^2}{M^2} + \frac{3\left( g^2+g'^2 \right)^2}{16\lambda^2} \frac{\tilde{M}^4}{M^4} \right) \left( 1-\frac{g^2+g'^2}{2\lambda} \frac{\tilde{M}^2}{M^2} \right)^{1/2} ~,
\end{align}
in the WKB regime when motion of $ \varphi $ is sufficiently slow in the time scale of the Higgs oscillation. As for the quark channels, they are open for both $ \varphi >0 $ and $ \varphi <0 $ regimes. Specifically, the top and bottom quarks are massless during $ \varphi <0 $ so the channels are always opened, while during $ \varphi >0 $ bottom quark channel is never forbidden due to the small Yukawa coupling $ y_b $ but top quark channel should satisfy 
\begin{align}
    \xi^2 >& \xi^2_c \left( 1-\frac{\lambda}{y^2_t} \right) \equiv \xi_t^2 \simeq 4351^2 ~.
\end{align}
The decay rates in the WKB regime are given as follows 
\begin{align}
	\Gamma_{h_0 \rightarrow \tilde{t} \bar{\tilde{t}}} =&\left\{
	\begin{array}{ll} 
	    \frac{3y^2_t}{16\pi} \left( 3\xi\alpha |\varphi| \right)^{1/2} M &, \varphi <0 \\ [0.3cm]
	    \frac{3y^2_t}{16\pi} \left( 6\xi\alpha \varphi \right)^{1/2} M \left( 1- \frac{y^2_t}{\lambda} \frac{\tilde{M}^2}{M^2} \right)^{3/2} &, \varphi >0
	\end{array}
	\right. \label{eq-Higgs-decay-top} \\
	\Gamma_{h_0 \rightarrow \tilde{b} \bar{\tilde{b}}} =&\left\{
	\begin{array}{ll} 
	    \frac{3y^2_b}{16\pi} \left( 3\xi\alpha |\varphi| \right)^{1/2} M &, \varphi <0 \\ [0.3cm]
	    \frac{3y^2_b}{16\pi} \left( 6\xi\alpha \varphi \right)^{1/2} M \left( 1- \frac{y^2_b}{\lambda} \frac{\tilde{M}^2}{M^2} \right)^{3/2} &, \varphi >0
	\end{array}
	\right. ~.
\end{align}
Note that the last line in Eq.~\eqref{eq-higgs-quarks-interaction} indicates that the Higgs particles are also able to decay into quarks, which will be discussed later. 

As for the scalaron decay, the dominant channel is decaying into Higgs particles. The decay is kinematically allowed only when $ \xi \ll 1 $ or $ \varphi $ becoming small when it gets close to origin or at very late time, as is shown by 
\begin{align}
    \Gamma_{\varphi \rightarrow \tilde{\delta h} \tilde{\delta h}} 
    &=\left\{ 
    \begin{array}{ll}
        \frac{3}{16\pi} \frac{M^3}{M^2_{\rm pl}} \left( \frac{1}{6} +\xi \right)^2 \left( 1- 12\xi \alpha |\varphi| \right)^{1/2} &, \varphi <0 \\ [0.3cm] 
        \frac{3}{16\pi} \frac{\tilde{M}^3}{M^2_{\rm pl}} \left( \frac{1}{6}-2\xi \frac{M^2}{\tilde{M}^2} \right)^2 \left( 1- 24 \xi \frac{M^2}{\tilde{M}^2} \alpha \varphi \right)^{1/2} &, \varphi >0 ~. \label{eq-scalaron-decay}
    \end{array}
    \right.
\end{align}
Therefore, the scalaron is expected to play an important role at late time in the determination of the final reheating temperature.

With the results above, we are now ready to discuss the effects of the perturbative decay on the whole reheating process in the mixed Higgs-$ R^2 $ model. 

\subsection{In the presence of tachyonic preheating}\label{Subsec-with-tachyonic}

In this subsection, we focus on the case where strong tachyonic instability will take place at the early stage of preheating and result in an early radiation-dominated era. In this case, the perturbative decay of the background fields will affect the tachyonic preheating in two ways as shown below. Firstly, the decay of the Higgs particles produced by tachyonic effect can reduce their backreaction strength on the homogeneous background, for example the rescattering effect considered in Ref.~\cite{Bezrukov:2020txg}. Secondly, even if the tachyonic preheating and the backreaction can transfer most of the energy from homogeneous background into relativistic particles, as long as there is some fraction of homogeneous field (mainly scalaron) remains, it will finally dominate the universe again as it is redshifted by the cosmic expansion in a slower manner than relativistic particles. As a result, perturbative decay eventually determines the reheating temperature regardless of the detail of preheating. 

\subsubsection{During tachyonic preheating}

First, we will discuss how perturbative decay affects the efficiency of the rescattering process between homogeneous and inhomogeneous modes after moderate tachyonic particle production. This is a quantum effect which is not considered in the classical lattice simulation in Ref.~\cite{Bezrukov:2020txg}. 

\begin{figure}[h]
	\centering
	\begin{subfigure}
		\centering
		\includegraphics[width=.45\textwidth]{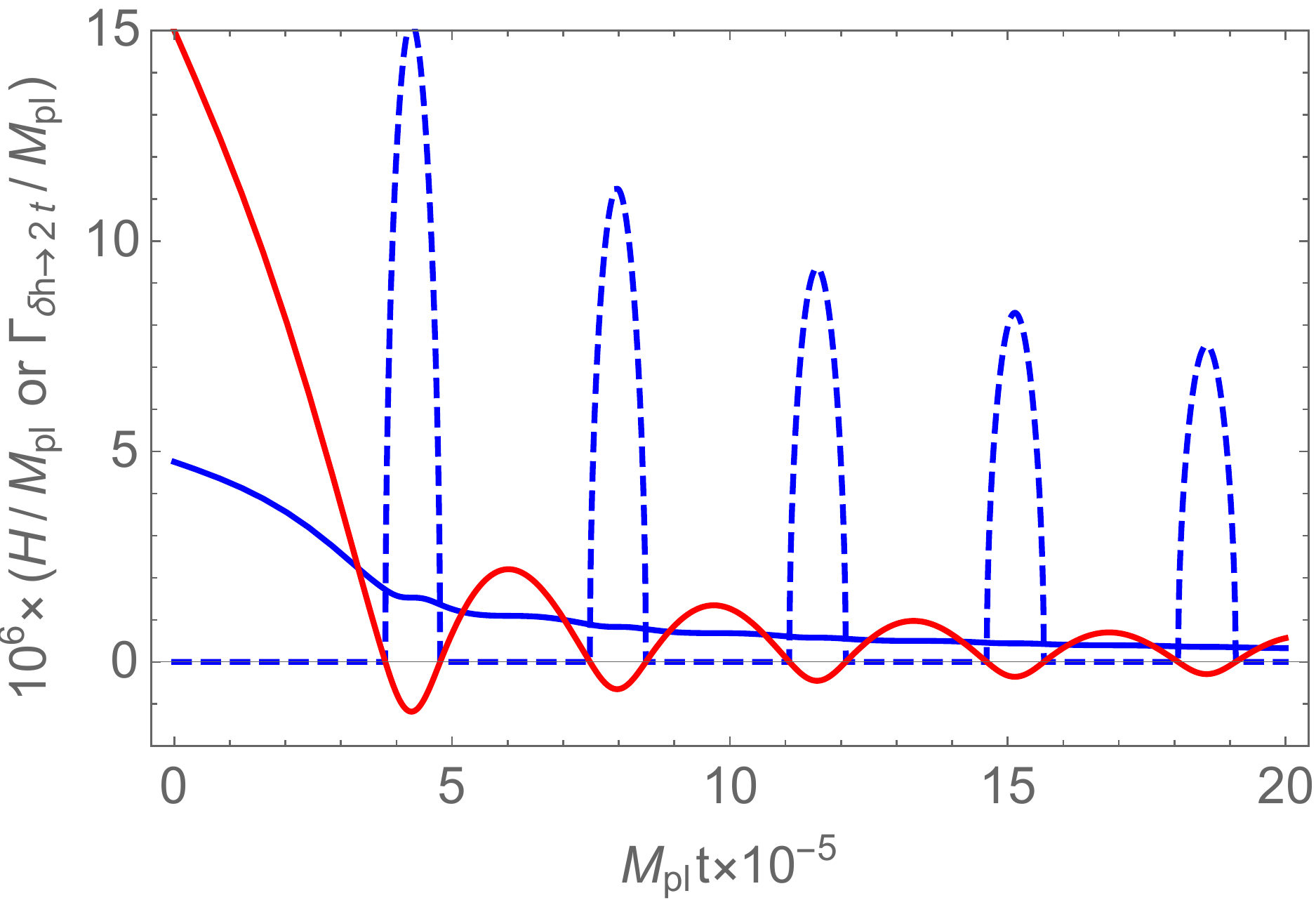}
	\end{subfigure}
	\begin{subfigure}
		\centering
		\includegraphics[width=.45\textwidth]{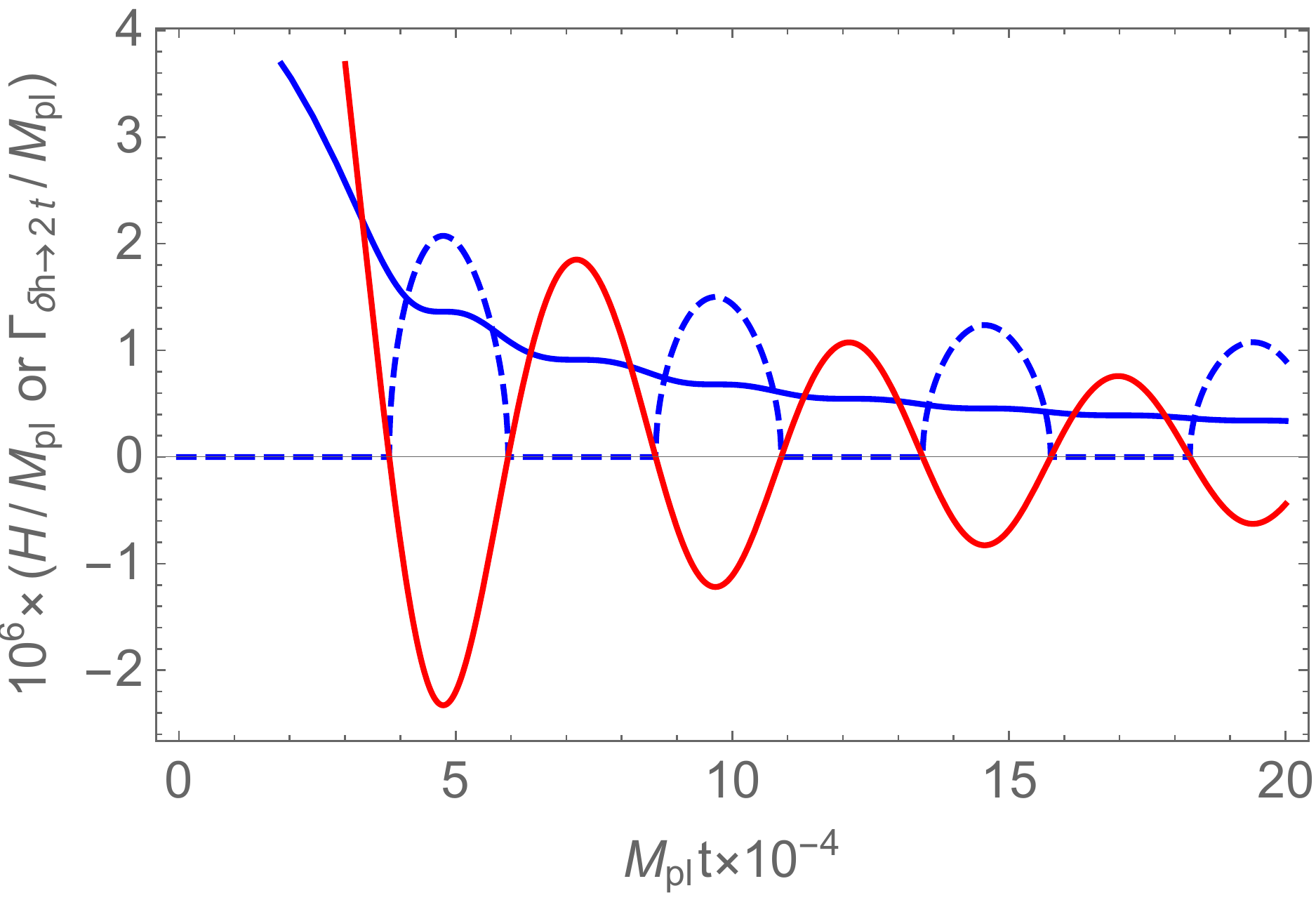}
	\end{subfigure}
	\caption{Comparison of Hubble scale $ H $ (blue, solid) and the decay rate $ \Gamma_{\delta h \rightarrow t \bar{t}} $ (blue, dashed). The evolution of the scalaron (red) is only for reference. Left: $ \xi =4000 $. Right: $ \xi =200 $. The decay rate vanishes when $ \varphi >0 $, because the top quarks gain large mass in these periods from the Higgs non-zero expectation value, which forbids the decay kinematically. Only for $ \xi $ comparable with $ \xi_s $ (which is not shown here), the this channel is allowed during $ \varphi >0 $, as mentioned in the previous subsection.}
	\label{fig-delta-Higgs-decay-rate}
\end{figure}

In the last line of Eq.~\eqref{eq-higgs-quarks-interaction}, one can see that the inhomogeneous Higgs can efficiently decay into top (and bottom) quarks as the homogeneous Higgs does, especially when $ \varphi <0 $. With similar treatment in Appendix~\ref{Appendix-decay-rate}, one can easily obtain the decay rate of this channel 
\begin{align}
    \Gamma_{\delta h \rightarrow t \bar{t}} =&\left\{
    \begin{array}{ll} 
        \frac{3y^2_t}{16\pi} \left( 3\xi\alpha |\varphi| \right)^{1/2} M &, \varphi <0 \\ [0.3cm]
        \frac{3y^2_t}{16\pi} \left( 6\xi\alpha \varphi \right)^{1/2} M \left( 1- \frac{y^2_t}{\lambda} \frac{\tilde{M}^2}{M^2} \right)^{3/2} &, \varphi >0 ~.
    \end{array}
    \right. 
\end{align}
In the derivation of the expressions above, we have neglected the cosmic expansion because the decay rate is much larger than the Hubble scale. For example, during the first scalaron oscillation, 
\begin{align}
	\Gamma_{\delta h \rightarrow t \bar{t}} = \frac{3y^2_t}{16\pi} \left( 3\xi\alpha |\varphi| \right)^{1/2} M \simeq & \frac{3y^2_t}{16\pi} \left( 3\xi C_1 \frac{M_c}{M} \right)^{1/2} M \\
	\simeq& 0.86 \times \left( \frac{M^2}{M^2_c} -1 \right)^{1/4} M_c > H \simeq 0.12 M_c
\end{align}
for $ \xi \gsim 100 $ on which we will focus in the following (see Fig.~\ref{fig-delta-Higgs-decay-rate} for two examples)\footnote{Possible reduction of the decay rate $ \Gamma_{\delta h \rightarrow t \bar{t}} $ is the Lorentz factor if the Higgs particles produced by tachyonic instability are relativistic. However, this depends on the momentum distribution of $ \delta h $ which is determined by the detail of the tachyonic preheating. This is beyond the scope of this paper. If most of the Higgs particles are indeed relativistic, then this decay channel may be neglected.}. 

\begin{figure}[h]
	\centering
	\begin{subfigure}
		\centering
		\includegraphics[width=.45\textwidth]{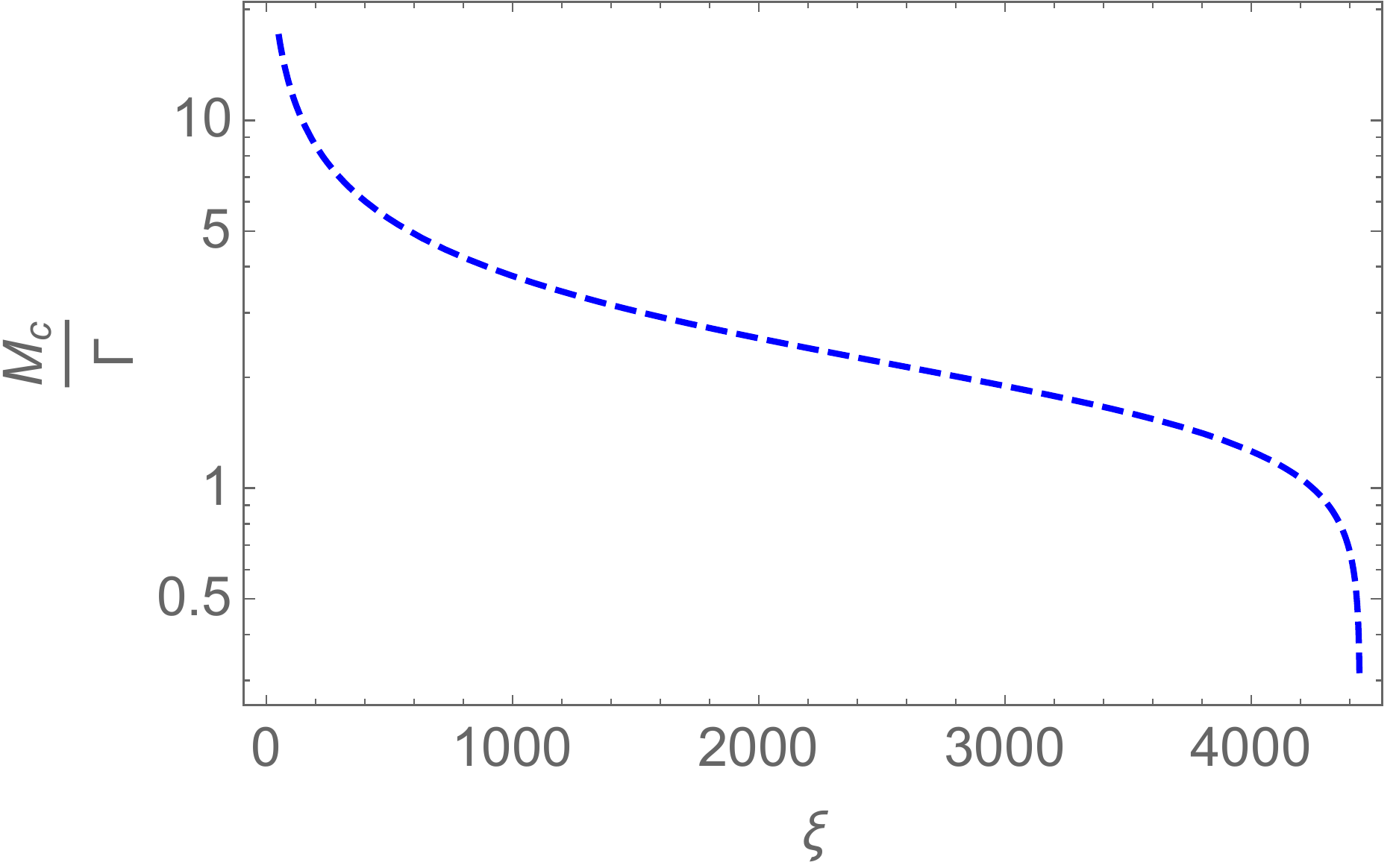}
	\end{subfigure}
	\begin{subfigure}
		\centering
		\includegraphics[width=.45\textwidth]{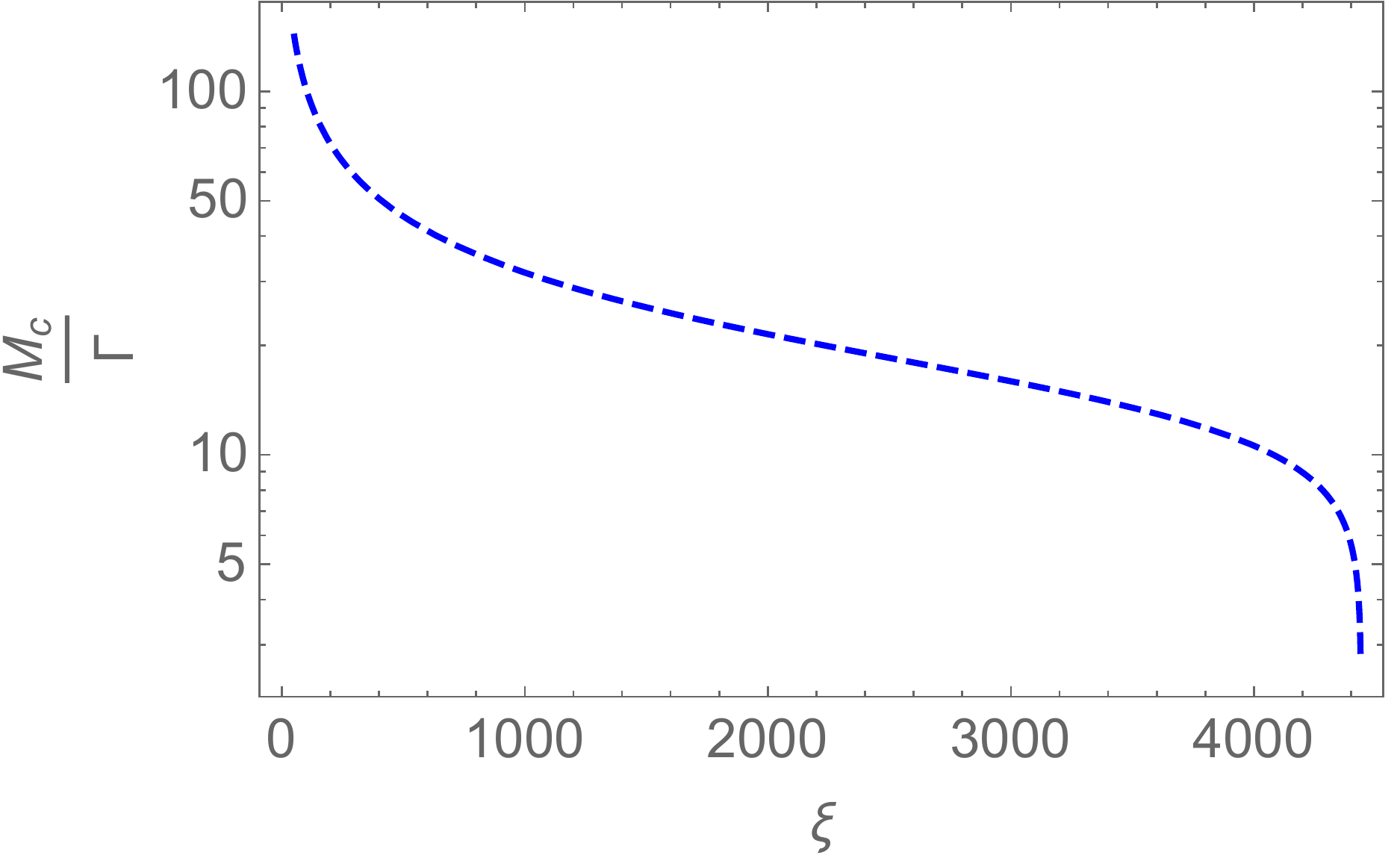}
	\end{subfigure}
	\caption{Decay time scale of the produced Higgs particles $ \delta h $ into top quarks. Eq. Left: $ C_1=0.25 $. Right: $ C_1=0.005 $.}
	\label{fig-delta-Higgs-decay-time -scale}
\end{figure}

Now, we consider the regime $ \varphi <0 $ only. The typical time scale of the decay $ \delta h \rightarrow t\bar{t} $ in the unit of $ M_c $ can be calculated as 
\begin{align}
	\frac{1}{M_ct}\sim M_c^{-1} \overline{\Gamma}_{\delta h \rightarrow t \bar{t}} =&\frac{3y^2_t}{16\pi} \overline{\left( 3\xi\alpha |\varphi| \right)^{1/2}} \frac{M}{M_c} \\
	\approx & \frac{3y^2_t}{16\pi} \sqrt{3 C_1}C_{m_h} \left(\frac{\lambda}{3}\right)^{1/4} \sqrt{\frac{M_{\rm pl}}{M_c}} \left(\frac{M^2}{M_c^2}-1\right)^{1/4} \label{eq-delta-Higgs-decay-time-scale}
\end{align}
where we have used Eq.~\eqref{eq-scalaron-varphi-1} as well as the result in Ref.~\cite{He:2020ivk} that the time-averaged $ \sqrt{|\varphi|} $ is captured by $ C_{m_h}=0.64 $. The result of Eq.~\eqref{eq-delta-Higgs-decay-time-scale} is shown in Fig.~\ref{fig-delta-Higgs-decay-time -scale}. Since $ C_1=0.25 $ is only valid for the first scalaron oscillation, we also consider $ C_1=0.005 $ to capture the decay of scalaron amplitude at late time. As can be seen, for most part of the parameter space, the time scale $ M_c \Gamma_{\delta h \rightarrow t \bar{t}}^{-1} $ is less than $ \sim 50 $ scalaron oscillations which is much shorter than the preheating time scale shown in the examples in Ref.~\cite{Bezrukov:2020txg}. Therefore, the number of Higgs particles that participate in the rescattering process will decrease considerably. 
Although the Higgs particles can decay into relativistic particles of other species, these decay products cannot efficiently interact with the homogeneous scalaron because they are coupled only through the conformal factor which is Planck-suppressed. Consequently, one may expect that the efficiency of the preheating process consisting of moderate tachyonic instability and rescattering can receive a non-negligible reduction due to the decay of the produced Higgs particles, which may also allow the perturbative decay process to dominate the last stage of reheating and eventually determine the reheating temperature $ T_r $. 

\subsubsection{After tachyonic preheating}

Next, we consider the case where a significant amount of energy has been already transferred from the homogeneous background fields to relativistic particles. It is interesting to study whether the homogeneous part can dominate over the radiation at late time, when the preheating processes can no longer significantly increase the energy fraction of the relativistic particles. If the answer is positive, it is also important to investigate how much the initial radiation energy affects the final reheating temperature that is determined by late-time perturbative decay of the homogeneous part. We address these questions with numerical methods. 

To determine the reheating temperature, one needs to find out the Hubble scale at the moment when reheating is completed. Here, by completion of reheating, we refer to the onset of the radiation-domination with no homogeneous fields that can dominate the universe again, which is actually ambiguous to define in a numerical calculation for the following reasons. Due to the accumulation of error, it is challenging to obtain reliable long-time numerical calculation to check whether the universe becomes matter-dominated again at late time. For example, it is usually too long for the numerical calculation to run for a scalaron decay time scale $ \Gamma_{\varphi}^{-1} $. Fortunately, in perturbative decay process considered here, the decay rate for scalaron is approximately constant at late time while the Hubble parameter $ H $ decays with time, so it is certain that the homogeneous background fields will be completely depleted at late time. Another reason is that the termination time chosen for the numerical calculation as the completion of reheating affects the Hubble scale (so the reheating temperature) due to the cosmic expansion. Too early (late) time will result in overestimation (underestimation) of the reheating temperature to some extent. In this paper, we simply use the ratio between the energy densities of radiation and homogeneous background fields $ \rho_{\rm rad}/\rho_{\rm bg} >20 $ as the criterion to make sure that the equation of state parameter $ w $ converges to $ 1/3 $ properly, which avoids the long-time numerical calculation and too much overestimation of $ T_r $. 

In our numerical calculation, we solve the equations of motion for the background fields with the decay rates taken into account as friction terms, the Boltzmann equation for the radiation energy density, and the Friedmann equation as follows 
\begin{align}
	&\ddot{\varphi} +\left( 3H +\Gamma_{\varphi \rightarrow \tilde{\delta h} \tilde{\delta h}} \right) \dot{\varphi} + \frac{\partial U}{\partial \varphi} + \frac{\alpha}{2} e^{-\alpha \varphi} \dot{h}^2 =0 \\
	&\ddot{h} +\left( 3H +\Gamma_{h_0 \rightarrow \tilde{t} \bar{\tilde{t}}} +\Gamma_{h_0 \rightarrow \tilde{b} \bar{\tilde{b}}} +\Gamma_{h_0 \rightarrow \tilde{W}\tilde{W}} +\Gamma_{h_0 \rightarrow \tilde{Z}\tilde{Z}} \right) \dot{h} + e^{\alpha \varphi}\frac{\partial U}{\partial h} -\alpha \dot{\varphi}\dot{h} =0 \\
	&\frac{d \rho_{\rm rad}}{dt} +4H\rho_{\rm rad} = \Gamma_{\varphi \rightarrow \tilde{\delta h} \tilde{\delta h}} \dot{\varphi}^2 + \left( \Gamma_{h_0 \rightarrow \tilde{t} \bar{\tilde{t}}} +\Gamma_{h_0 \rightarrow \tilde{b} \bar{\tilde{b}}} +\Gamma_{h_0 \rightarrow \tilde{W}\tilde{W}} +\Gamma_{h_0 \rightarrow \tilde{Z}\tilde{Z}} \right) \dot{h}^2 \\
	& 3M^2_{\rm pl} H^2 =\rho_{\rm bg} +\rho_{\rm rad}
\end{align}
where $ \rho_{\rm rad} $ is the energy density of radiation while $ \rho_{\rm bg} $ is the energy density of the background fields given as 
\begin{align}
	\rho_{\rm bg} = \frac{1}{2} \dot{\varphi}^2 + \frac{1}{2} e^{-\alpha \varphi} \dot{h}^2 +U(\varphi, h) ~.
\end{align}
Also, we use step functions to impose the kinematic constraint for each decay channel. We use the numerical solution to estimate the equation of state parameter $ w $ and the reheating temperature $ T_r $ which are calculated as 
\begin{align}
    w&=-1-\frac{2}{3} \frac{\dot{H}}{H^2} \\
    T_r &= \left(\frac{90}{g_{\rm r} \pi^2}\right)^{1/4} \sqrt{H_{\rm r} M_{\rm pl}}
\end{align}
where $ g_{\rm r} = 106.75 $ and $ H_{\rm r} $ are the effective number of relativistic species and Hubble parameter at the end of reheating. We also define a temperature $ T_{\varphi} $ as a reference by assuming that Higgs becomes light enough so that the decay channel $ \varphi \rightarrow \tilde{\delta h} \tilde{\delta h} $ is always open 
\begin{align}
	T_{\varphi} \equiv& 0.2 \sqrt{M_{\rm pl} \overline{\Gamma}_{\varphi}} \nonumber \\
	\approx& 0.2 \sqrt{M_{\rm pl} } \left[ \left( \frac{\pi}{M} +\frac{\pi}{\tilde{M}} \right)^{-1} \left( \frac{3}{16\pi} \frac{M^3}{M^2_{\rm pl}} \left( \frac{1}{6} +\xi \right)^2 \frac{\pi}{M} + \frac{3}{16\pi} \frac{\tilde{M}^3}{M^2_{\rm pl}} \left( \frac{1}{6}-2\xi \frac{M^2}{\tilde{M}^2} \right)^2 \frac{\pi}{\tilde{M}} \right) \right]^{1/2} \nonumber \\
	=& 0.2 \sqrt{M_{\rm pl} } \sqrt{ \frac{3}{16\pi} \frac{M^3\tilde{M}}{M_{\rm pl}^2 (M+\tilde{M})} } \left[ \left( \frac{1}{6} +\xi \right)^2 + \frac{\tilde{M}^2}{M^2} \left( \frac{1}{6} -2\xi \frac{M^2}{\tilde{M}^2} \right)^2 \right]^{1/2} \label{eq-scalaron-only-temperature}
\end{align}
which returns to the standard result (see e.g. \cite{Gorbunov:2010bn,Gorbunov:2012ns}) when $ \xi \ll 1 $. 

\begin{figure}[h]
	\centering
	\begin{subfigure}
		\centering
		\includegraphics[width=.45\textwidth]{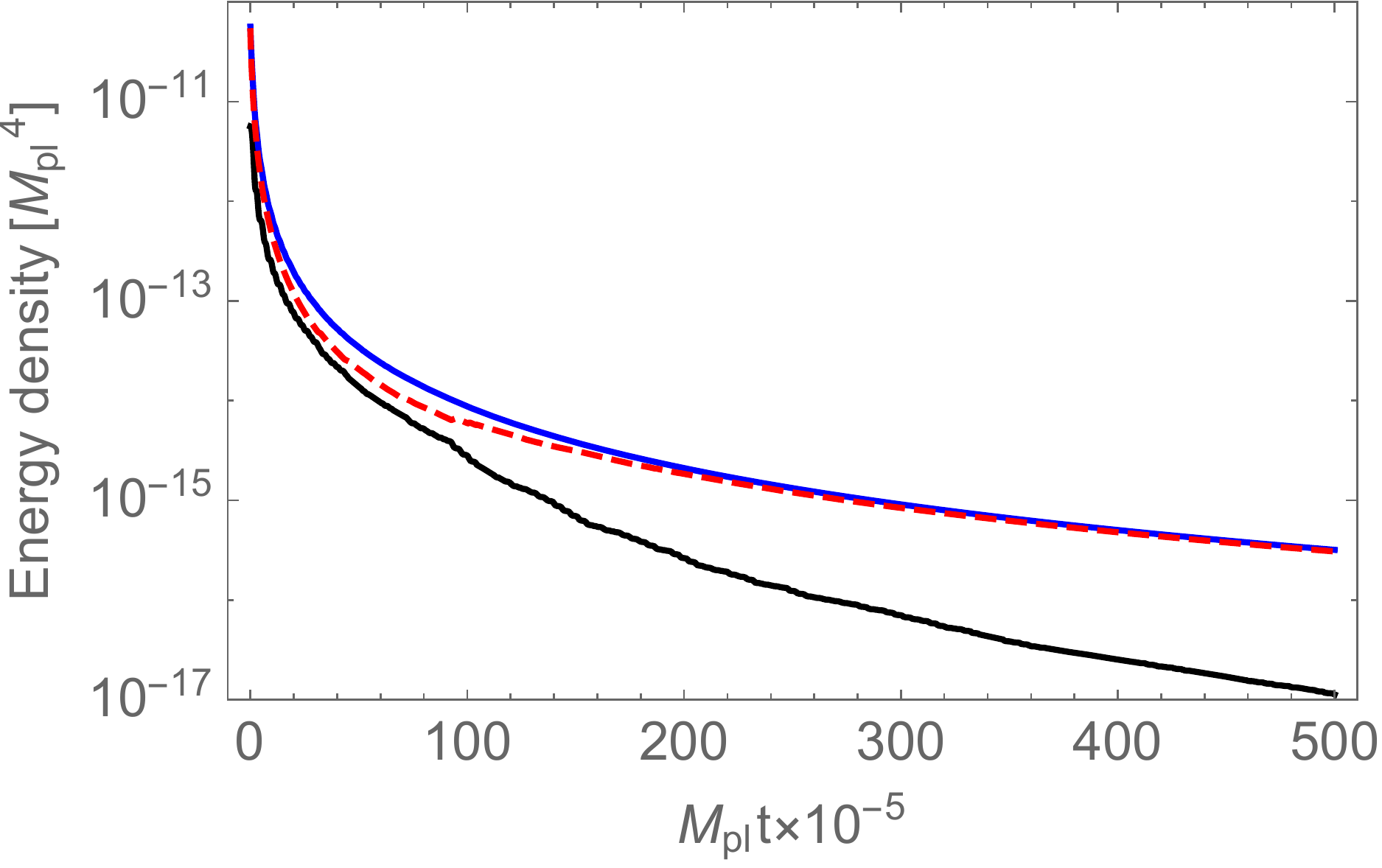}
	\end{subfigure}
	\begin{subfigure}
		\centering
		\includegraphics[width=.45\textwidth]{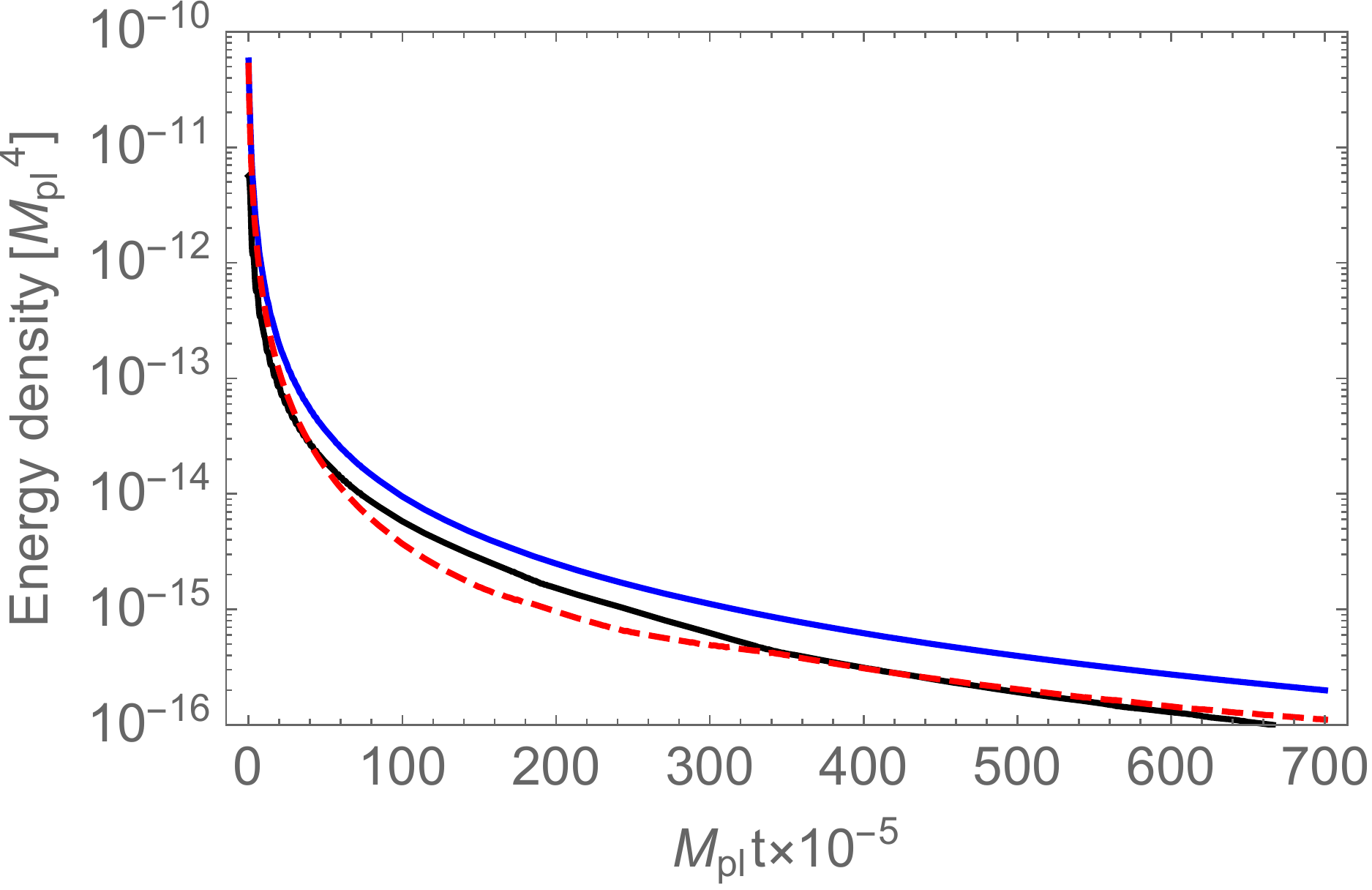}
	\end{subfigure}
    \begin{subfigure}
	    \centering
	    \includegraphics[width=.45\textwidth]{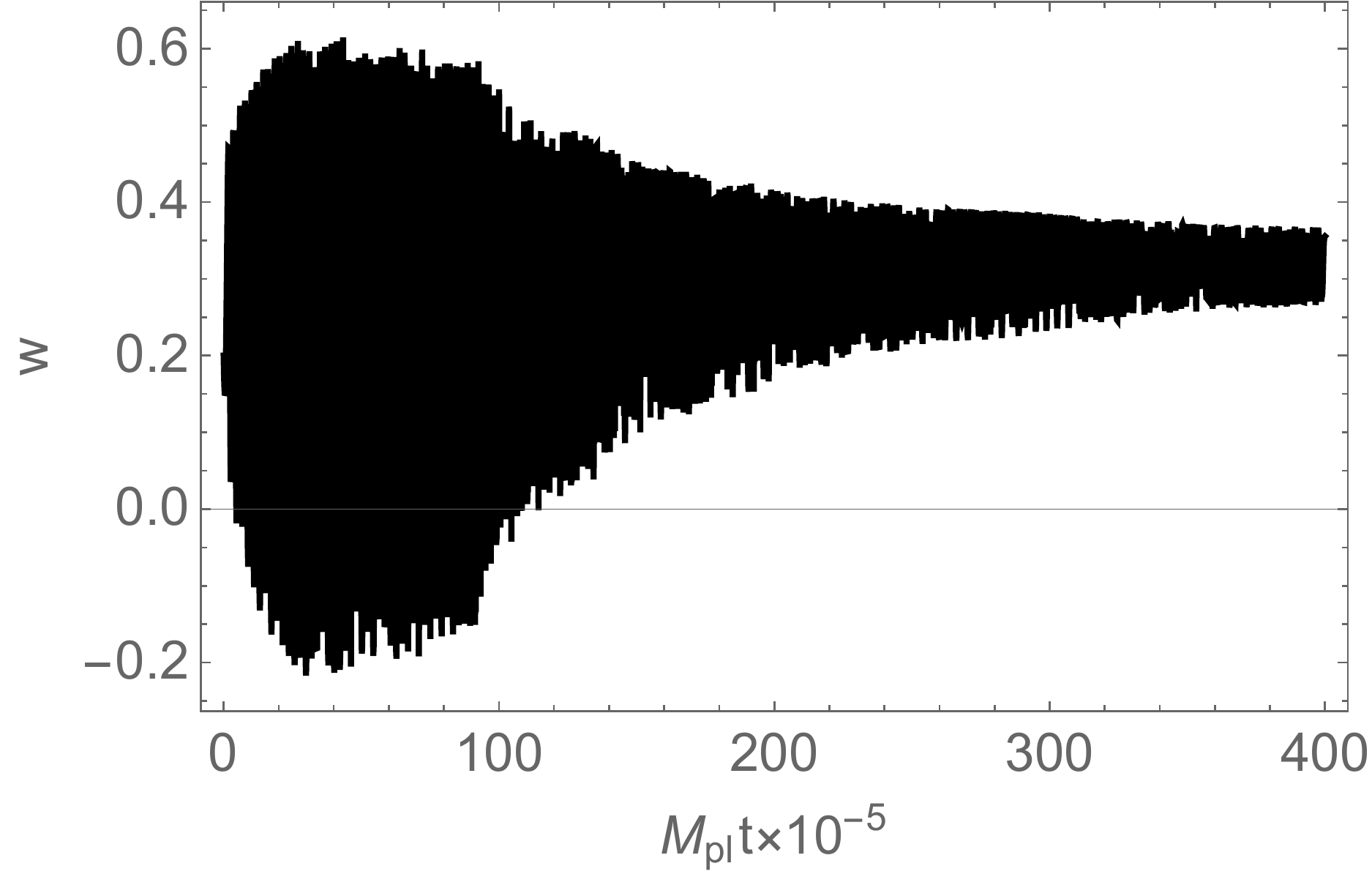}
    \end{subfigure}
    \begin{subfigure}
    	\centering
    	\includegraphics[width=.45\textwidth]{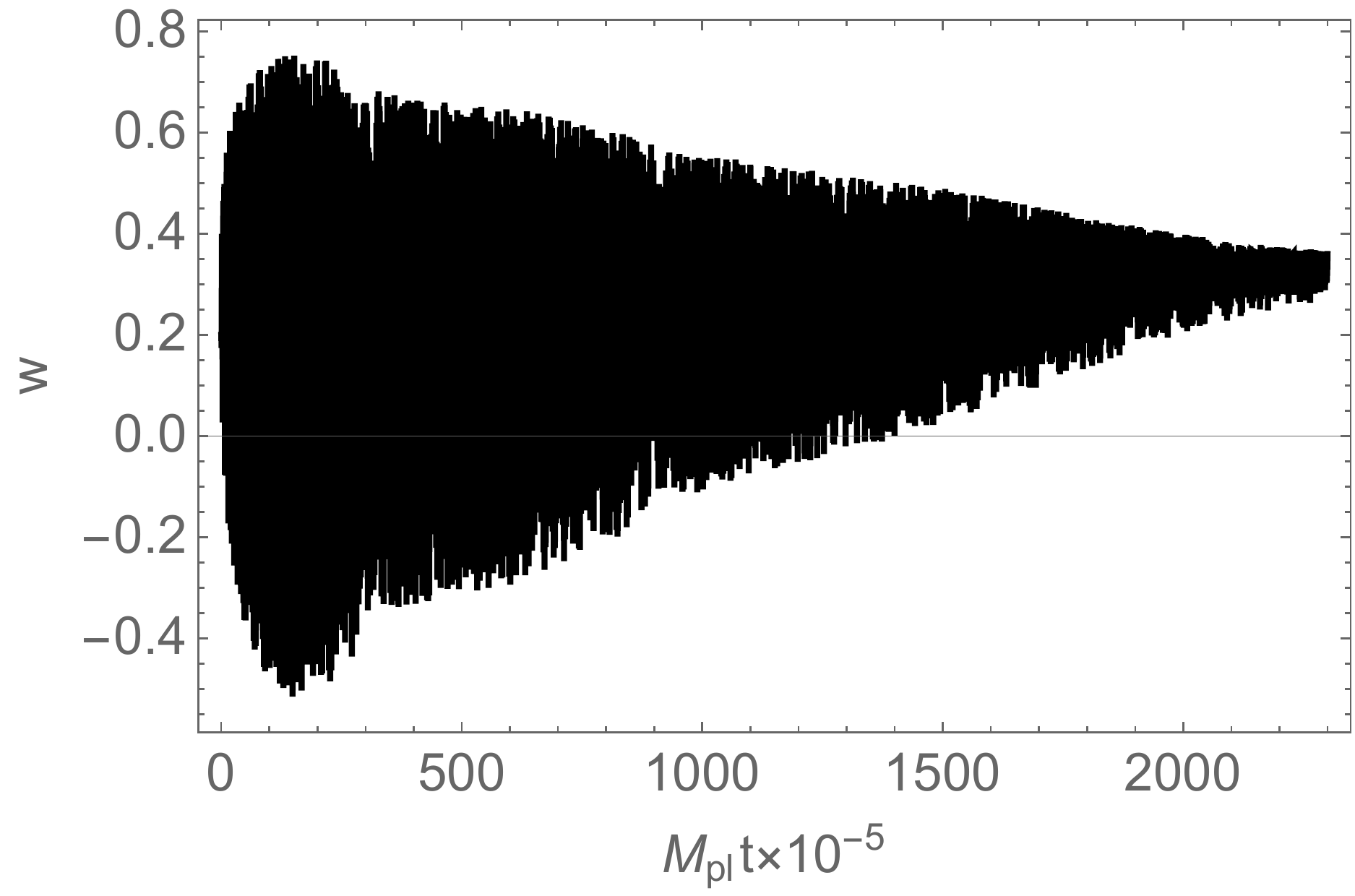}
    \end{subfigure}
	\caption{Up: Two examples of the evolution of $ \rho_{\rm rad} $ (red dashed), $ \rho_{\rm bg} $ (black solid), and $ 3M^2_{\rm pl} H^2 $ (blue solid) with $ \xi =3000 $ (left) and $ \xi=1000 $ (right) and initial radiation $ \rho_{\rm rad,ini} =9 \rho_{\rm bg,ini} =90 \% \times 3 M^2_{\rm pl} H^2_e $. Down: Two examples of the evolution of $ w $ up to $ \rho_{\rm rad}/\rho_{\rm bg} \gsim 20 $ with the same parameters as their upper panels respectively. In order to make the intersections clearer in upper right panel, we only show the evolution up to $ M_{\rm pl} t =7\times 10^7 $.}
	\label{fig-inirad-energy-w}
\end{figure}

Since we are considering a preexisting preheating process that has transferred a large amount of energy from $ \rho_{\rm bg} $ to $ \rho_{\rm rad} $, we set the initial conditions for them to be $ \rho_{\rm rad,ini} =9 \rho_{\rm bg,ini} =90 \% \times 3 M^2_{\rm pl} H^2_e $ where we take $ H_e $ to be the value at the end of inflation. Before going to the reheating temperature, we first show two examples of the evolution of $ \rho_{\rm rad} $, $ \rho_{\rm bg} $, $ 3M^2_{\rm pl} H^2 $, and $ w $ in Fig.~\ref{fig-inirad-energy-w}.
One can see that for larger $ \xi $, the homogeneous part does not really dominate over but at most becomes comparable to the radiation, while for smaller $ \xi $ the energy density of the homogeneous part indeed becomes larger than that of the radiation and we can obviously see two intersections between them in the upper right panel of Fig.~\ref{fig-inirad-energy-w}. This is reasonable because for larger $ \xi $ more decay channels are allowed and decay rates themselves also become larger. However, if we do not take perturbative decays into account, the energy density of the radiation will simply decay away quickly and the universe will be dominated by the homogeneous background field again. Also, it can be seen that $ w $ effectively converges to $ 1/3 $ when $ \rho_{\rm rad}/\rho_{\rm bg} \gsim 20 $, which shows that this criterion is reasonable. 

\begin{figure}[h]
	\centering
	\begin{subfigure}
		\centering
		\includegraphics[width=.45\textwidth]{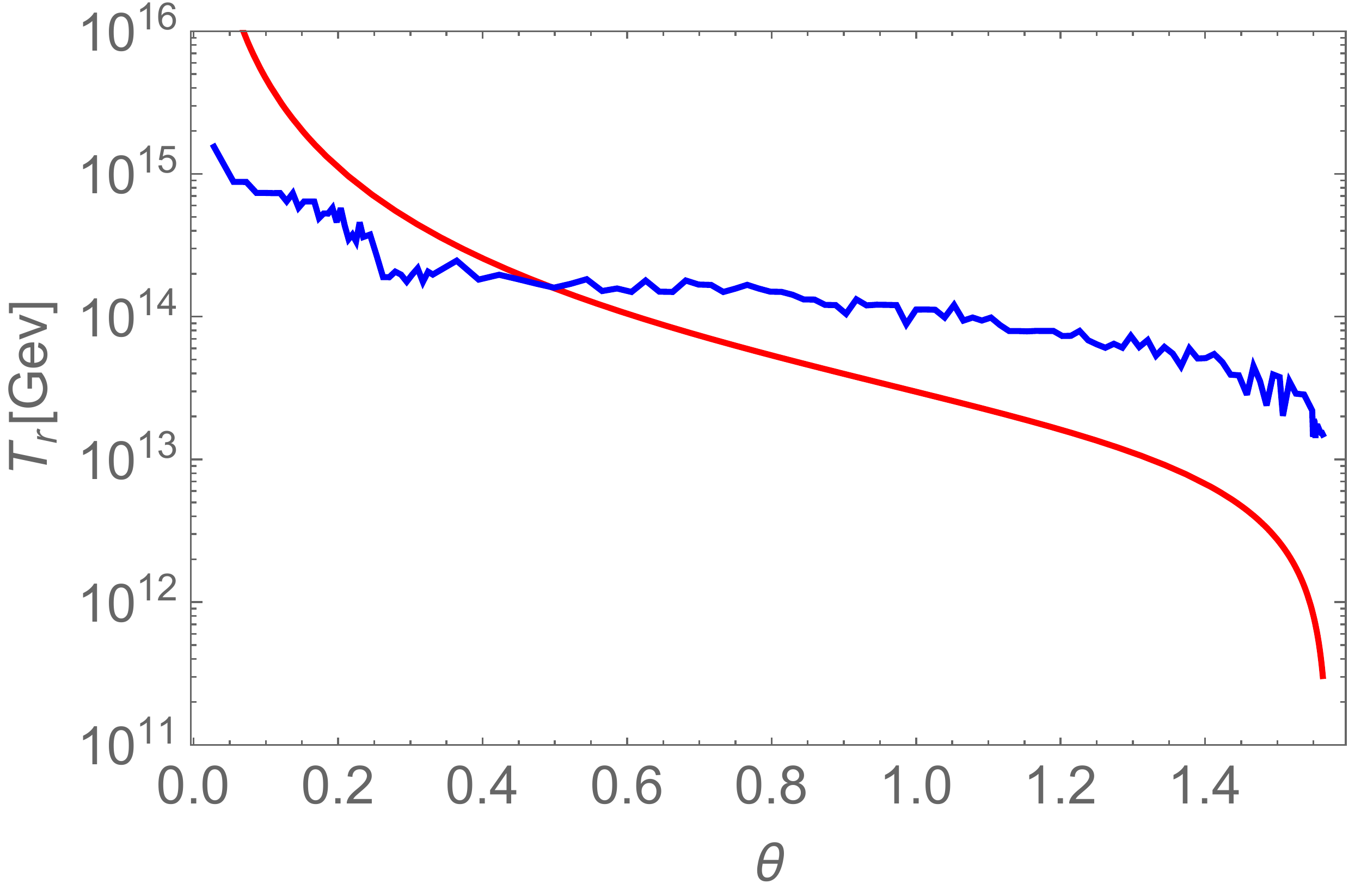}
	\end{subfigure}
	\begin{subfigure}
		\centering
		\includegraphics[width=.45\textwidth]{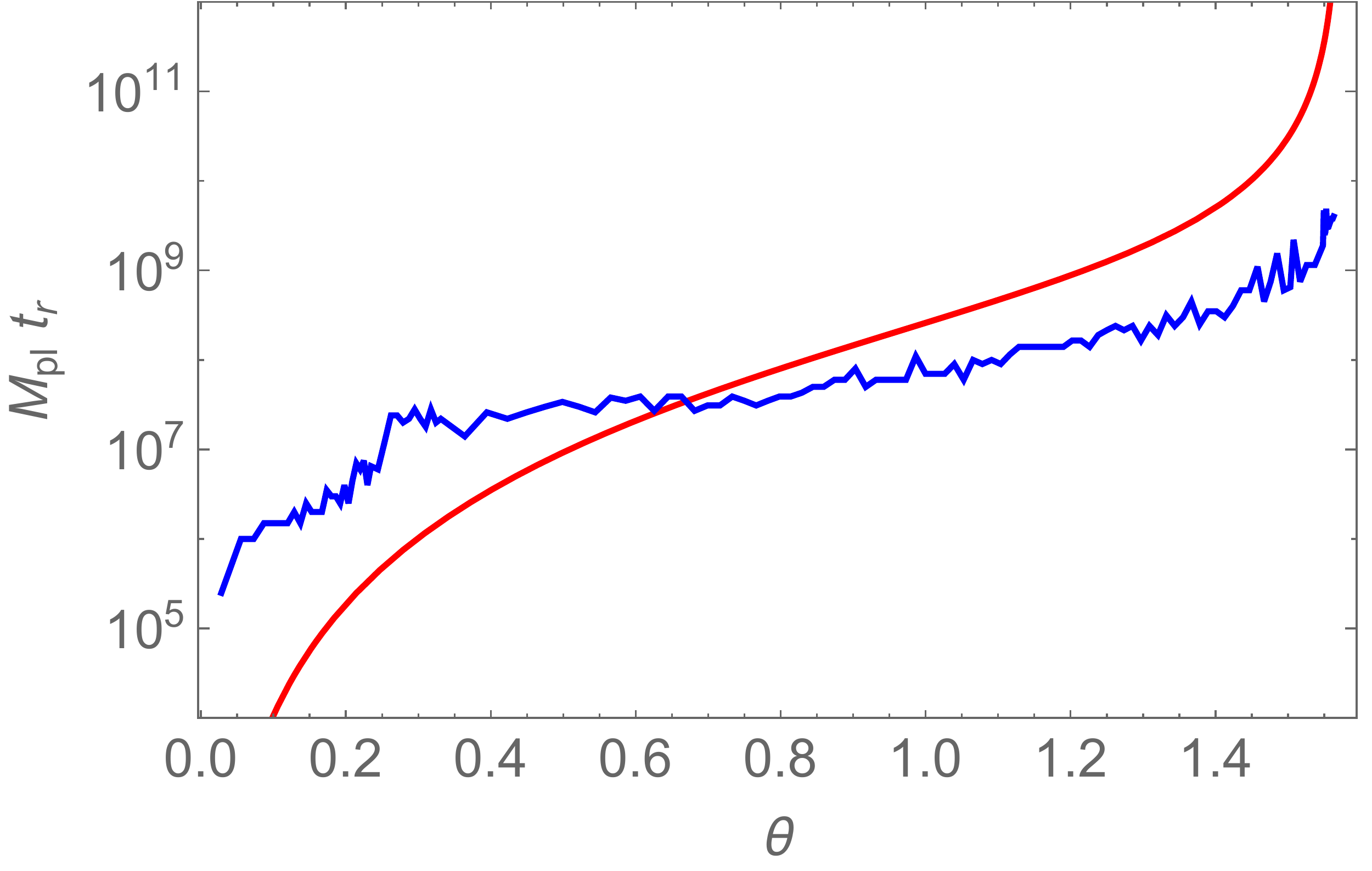}
	\end{subfigure}
	\caption{Left: reheating temperature $ T_r $ (blue) and the temperature calculated with only averaged scalaron decay rate $ T_{\varphi} $ (red) given in Eq.~\eqref{eq-scalaron-only-temperature}. Right: duration of perturbative reheating $ t_r $ (blue) and the scalaron decay time $ \Gamma_{\varphi}^{-1} $ (red). Both $ T_r $ and $ t_r $ are determined at the moment when $ \rho_{\rm bg} $ becomes smaller than $ 5 \% $ of $ \rho_{\rm rad} $. Here we use $ \theta $ as the model parameter.}
	\label{fig-inirad-temperature-duration}
\end{figure}

\begin{figure}[t]
	\centering
	\begin{subfigure}
		\centering
		\includegraphics[width=.32\textwidth]{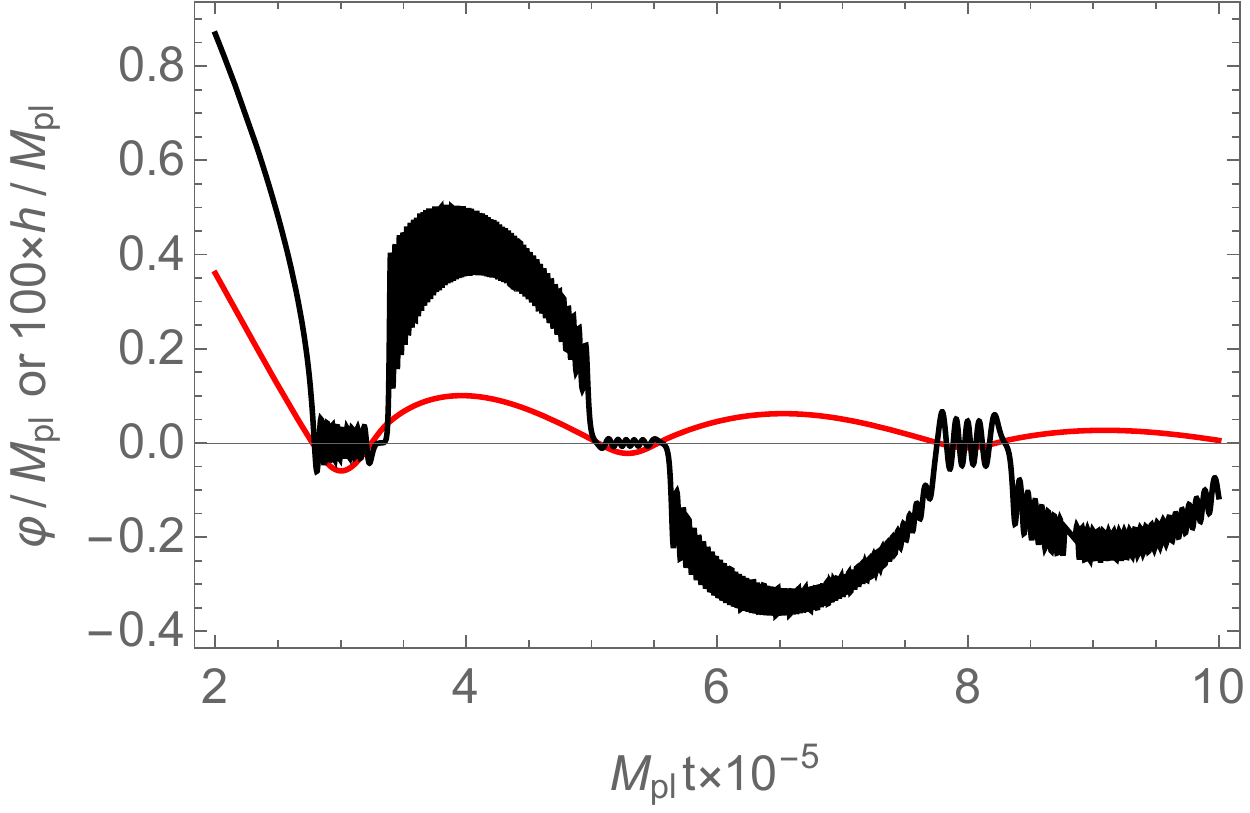}
	\end{subfigure}
	\begin{subfigure}
		\centering
		\includegraphics[width=.32\textwidth]{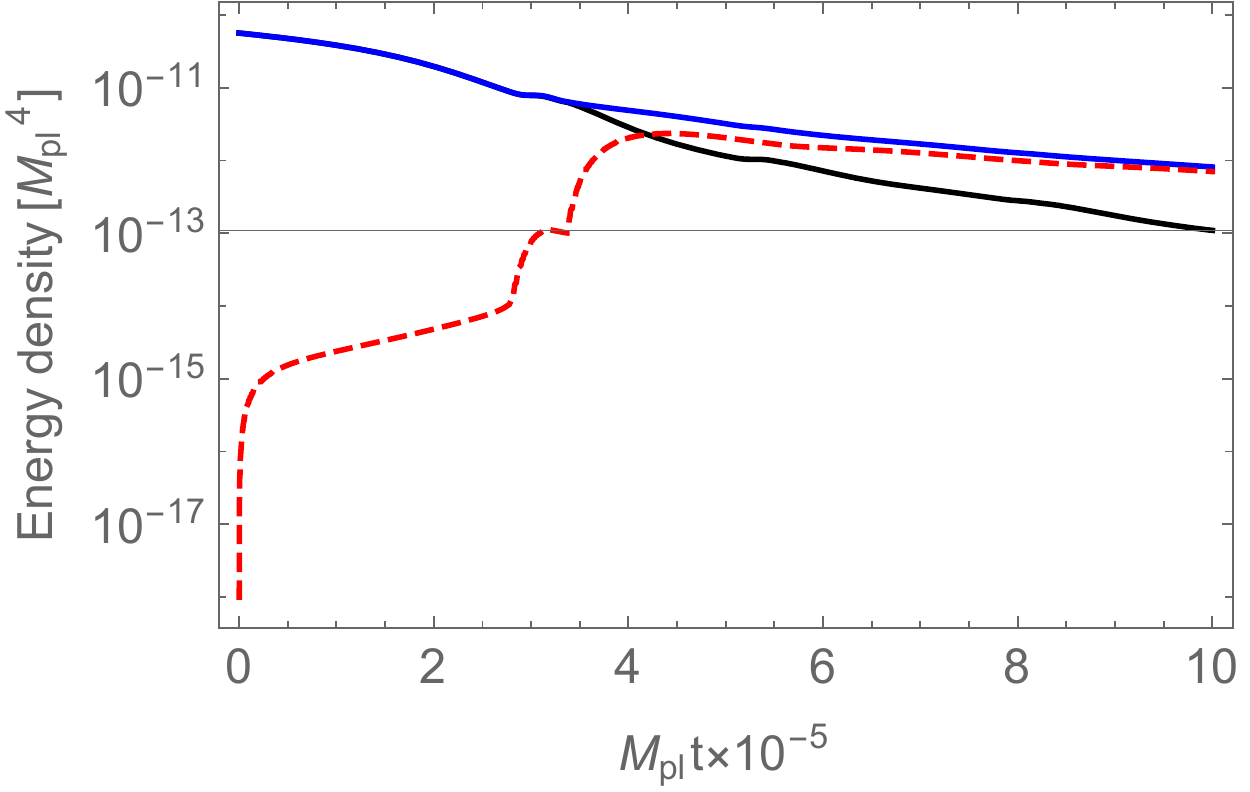}
	\end{subfigure}
    \begin{subfigure}
    	\centering
    	\includegraphics[width=.32\textwidth]{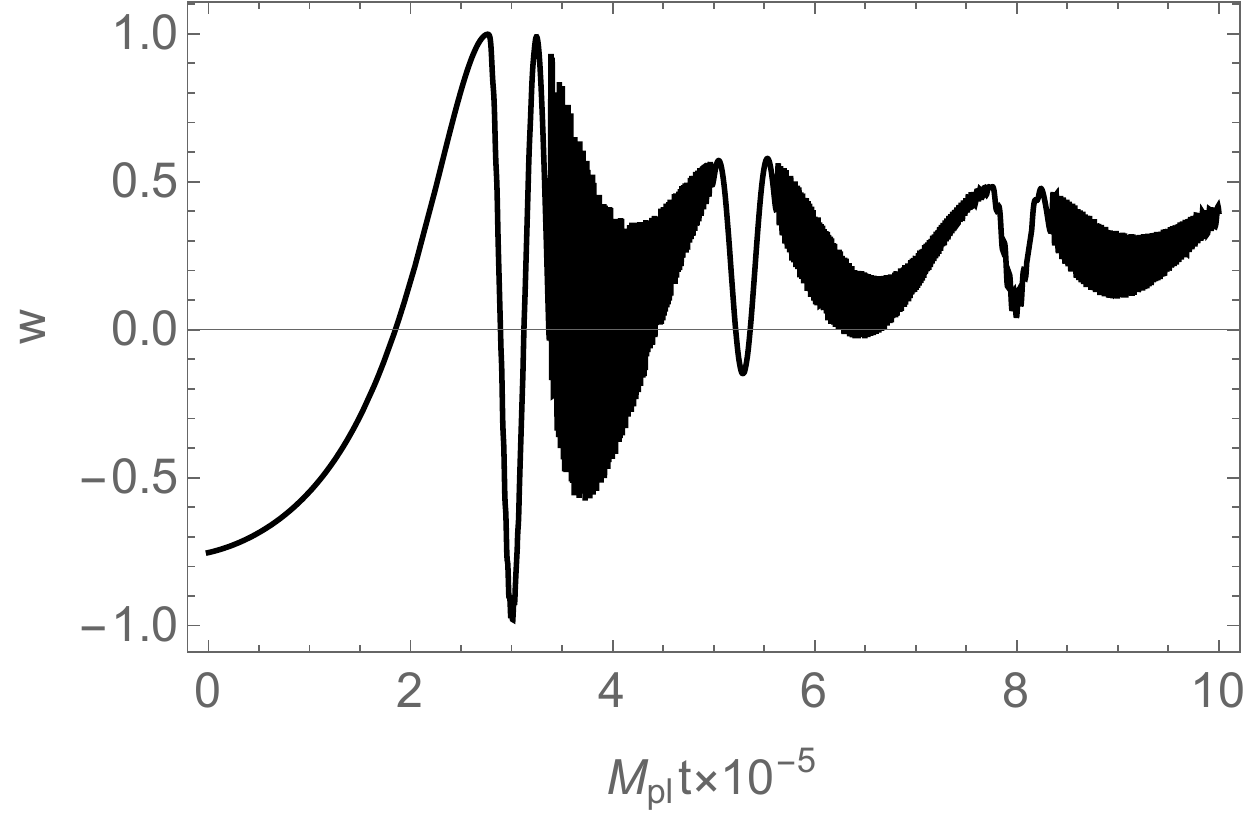}
    \end{subfigure}
    \begin{subfigure}
    	\centering
    	\includegraphics[width=.32\textwidth]{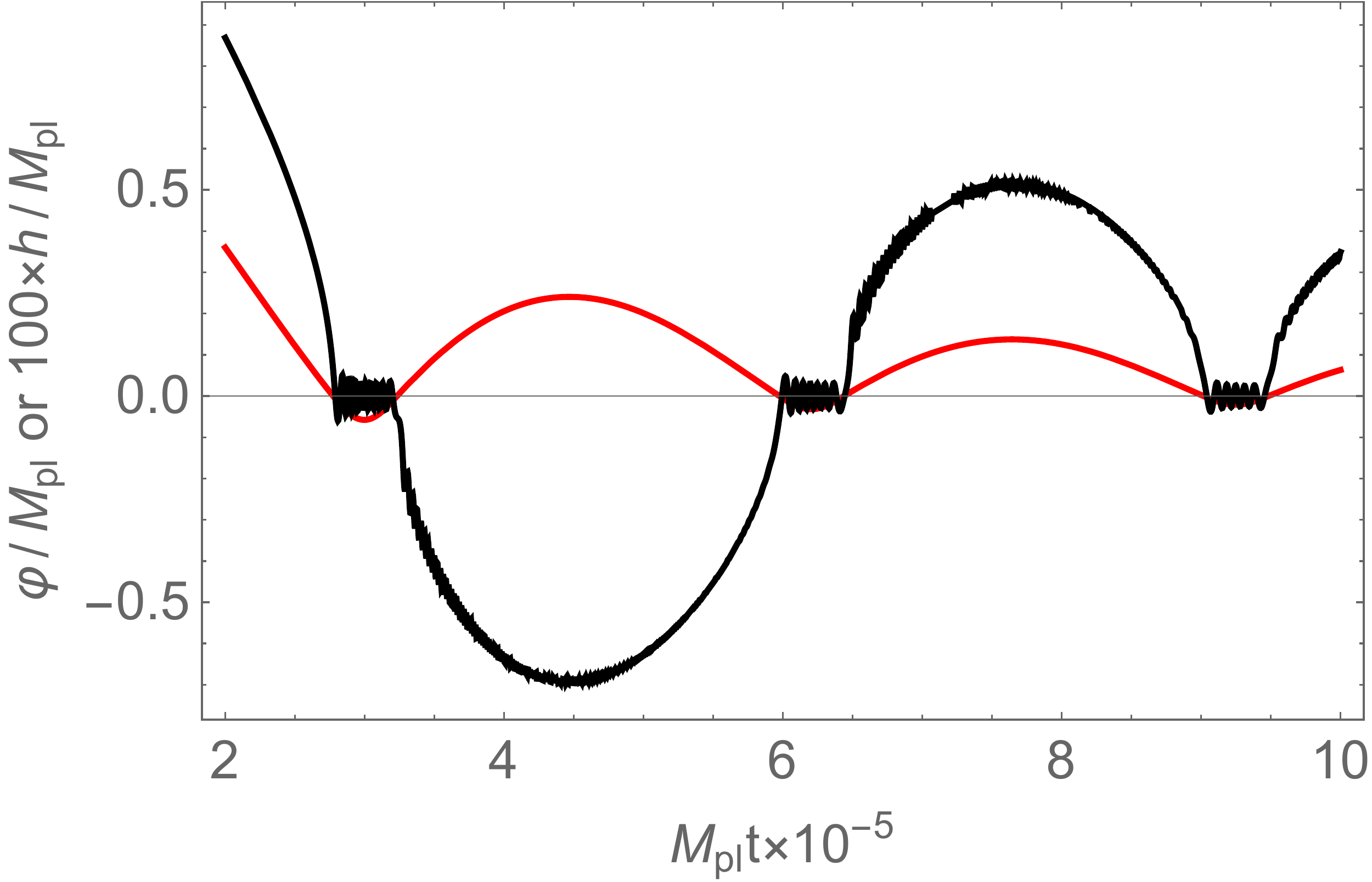}
    \end{subfigure}
    \begin{subfigure}
    	\centering
    	\includegraphics[width=.32\textwidth]{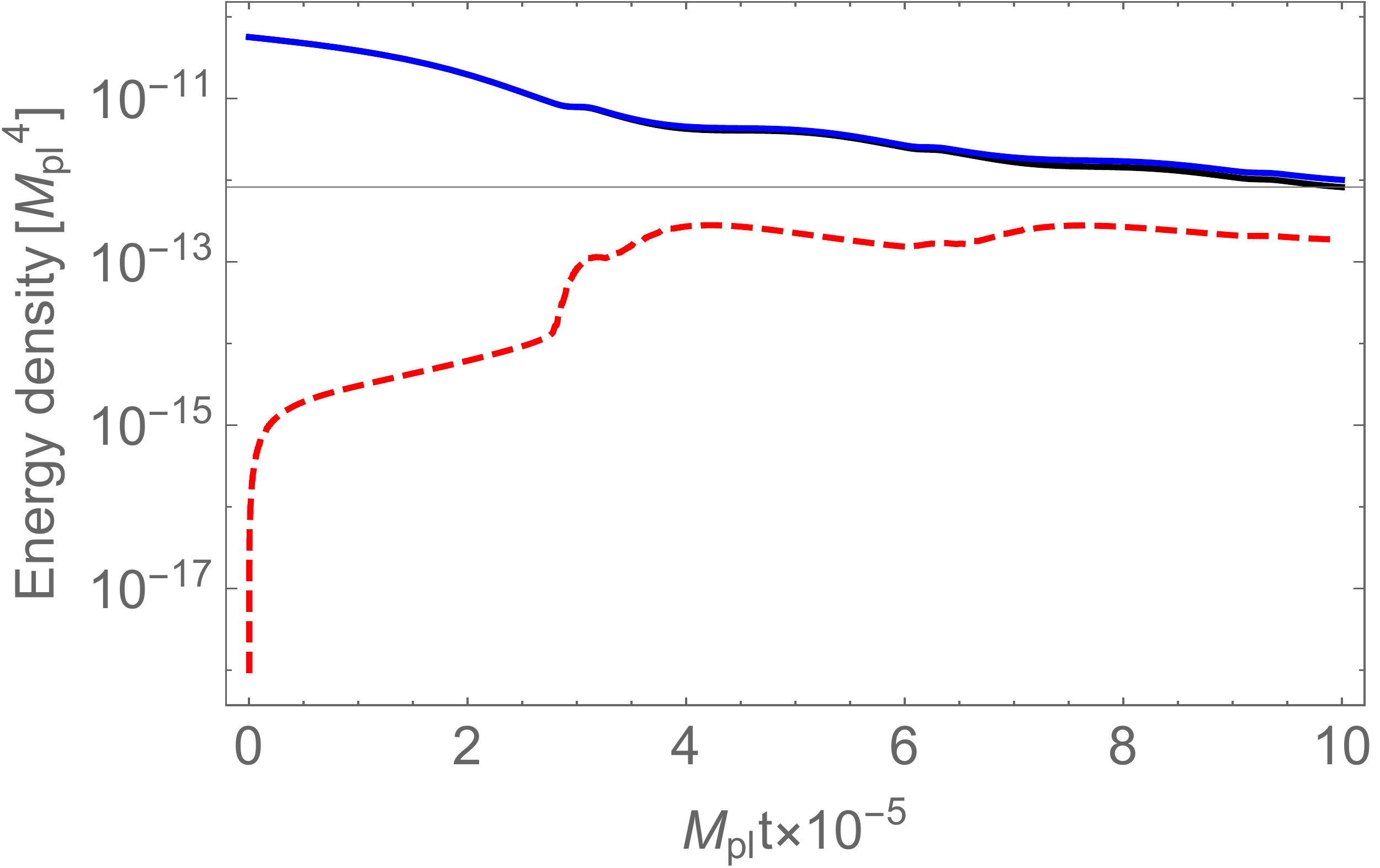}
    \end{subfigure}
    \begin{subfigure}
    	\centering
    	\includegraphics[width=.32\textwidth]{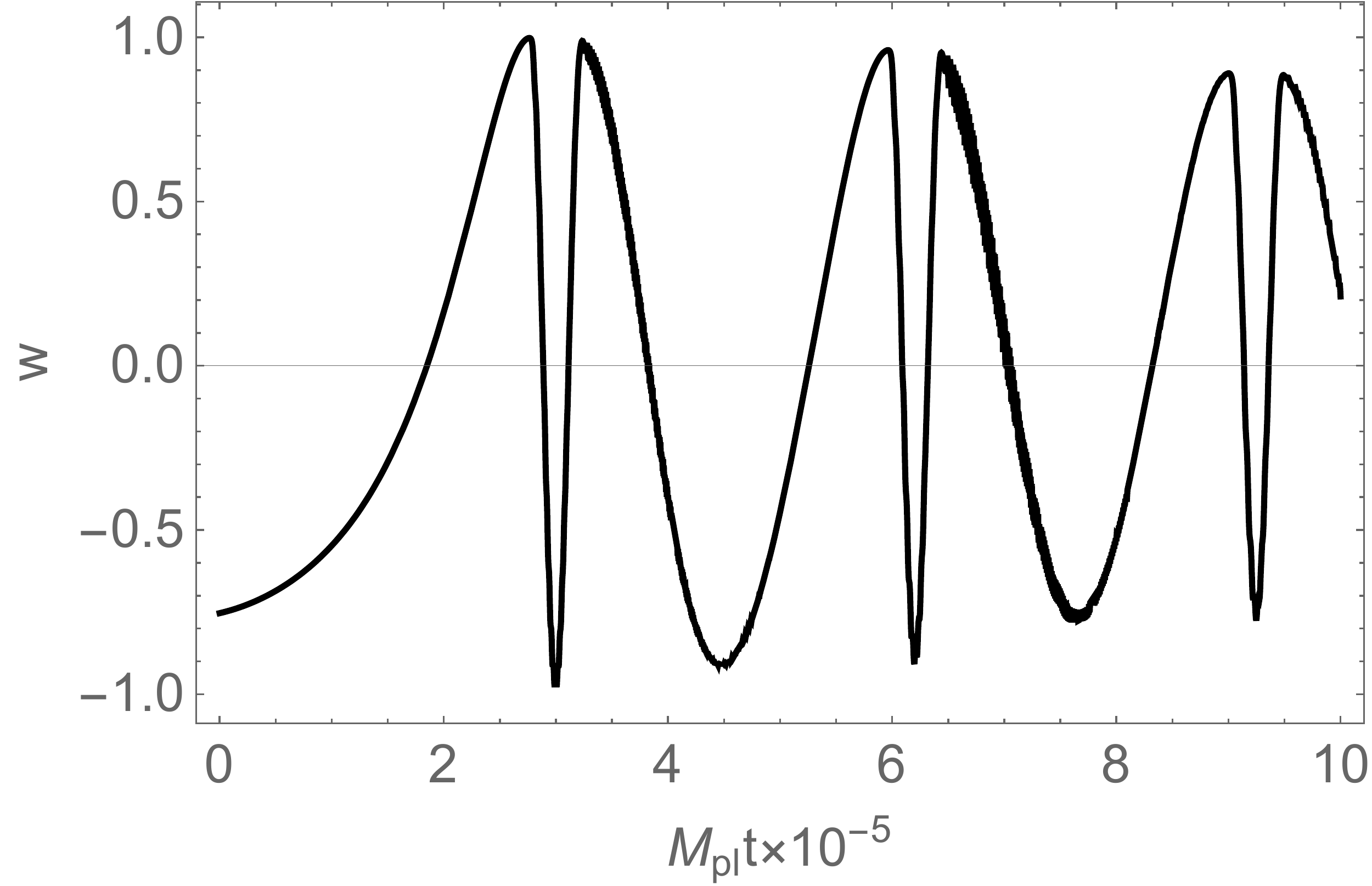}
    \end{subfigure}
	\caption{Up: nearly-critical parameter denoted as $ \xi_{nc} $. Down: far-from-critical parameter denoted as $ \xi_{fc} \simeq 1.001 \xi_{nc} $. Left: evolution of homogeneous scalaron (red) and Higgs (blue). Middle: $ \rho_{\rm rad} $ (red dashed), $ \rho_{\rm bg} $ (black solid), and $ 3M^2_{\rm pl} H^2 $ (blue solid). Right: $ w $. There is no initial radiation energy imposed for simplicity}
	\label{fig-illustrative-energy-distribution}
\end{figure}

Finally, we will discuss the reheating temperature for parameters $ \xi \gg \mathcal{O}(10) $ in this setup. We scan the parameter space $ 35 \leq \xi \leq \xi_s $ with adaptive step length and calculate the reheating temperature $ T_r $ defined at the moment when $ \rho_{\rm rad}/\rho_{\rm bg} $ exceeds 20 as well as the duration of this perturbative reheating $ t_r $. We do not further investigate the cases $ \xi <35 $ because it takes too long time for numerical calculation. From our results presented below, we expect that the reheating temperature quickly approaches to the $ R^2 $-inflation case as $ \xi $ decreases. The result is shown in Fig.~\ref{fig-inirad-temperature-duration} where we have used $ \theta $ as the model parameter because it treats the Higgs-like and $ R^2 $-like regions ``equally" (see Ref.~\cite{He:2020ivk} for more details.). This is one of the main results in this paper. 
One can see that once $ \theta $ approaches to the unitary bound $ \theta(\xi=\xi_s)\approx 0 $ (or $ \pi/2 $), the reheating temperature rapidly increases (or decreases) to reach the Higgs-limit (or $ R^2 $-limit), respectively. Similar argument also applies to the perturbative reheating duration. As for the comparison with $ T_{\varphi} $ (red line), we can see that $ T_r > T_{\varphi} $ for most part of the parameter space because the Higgs decay plays an important role in lifting up the reheating temperature. However, when $ \theta $ approaches the unitary bound in Higgs limit, $ T_{\varphi} $ becomes larger than $ T_r $, which is understandable. It is simply because we have assumed that the scalaron decay channel is always open, which is obviously NOT the realistic case because the large Higgs mass forbids the decay to occur until late time. Similar argument applies to the reheating time. Apart from these unrealistic part, we expect that if we wait for long enough time, the numerical results in blue will get closer to but not coincide with the red lines in Fig.~\ref{fig-inirad-temperature-duration}. Again, the reason is that the decay of Higgs is lifting up the reheating temperature and shorten the reheating time. More importantly, the small dips and peaks on the numerical results are physical instead of pure artificial due to the error of numerical calculation. As briefly mentioned in Sec.~\ref{Subsec-bg}, the cause of these small features is the $ \theta $-dependent energy distribution between scalaron and Higgs discussed in Appendix~\ref{Appendix-kinetic-Higgs}. If the model parameter is close to some critical point which realizes maximal tachyonic instability in the first scalaron ocsillation, more energy is stored in the Higgs field so that the effective decay rate of the whole system becomes larger because the decay rate of Higgs is larger than that of scalaron for most parameters. We show illustrative examples in Fig.~\ref{fig-illustrative-energy-distribution} to make the argument more convincing. 
We consider two cases (1) nearly-critical $ \xi_{nc} $ with larger oscillation amplitude of Higgs, and (2) far-from-critical $ \xi_{fc} \simeq 1.001 \xi_{nc} $ with smaller oscillation amplitude of Higgs, and there is no initial radiation imposed for simplicity. Since the two parameters are so close, the decay rates given in Sec.~\ref{Subsec-decay-rate} are almost the same for both cases. However, as can be easily seen in the comparison in Fig.~\ref{fig-illustrative-energy-distribution}, the nearly-critical parameter case induces much more efficient reheating process than the latter even with slightly smaller decay rates superficially, which can be recognized from the growth rate of $ \rho_{\rm rad} $ and the converging rate of $ w $ to $ 1/3 $. On the other hand, if the parameter is far from critical values, the energy in Higgs can be very small so that the system is effectively decaying only through $ \Gamma_{\varphi\rightarrow \tilde{\delta h}\tilde{\delta h}} $ until the inflaton encounters a chance to transfer the energy from scalaron to Higgs. However, the chance is expected to be quite rare if the necessary degree of fine-tuning for subsequent scalaron oscillations to be critical can be analyzed in a similar way as the first scalaron oscillation in Ref.~\cite{He:2020ivk}, and the corresponding effect would be smaller. Therefore, if the time resolution in the numerical calculation is high enough and the parameter scanning is done more finely, one would expect that the dips on the blue line in the left panel of Fig.~\ref{fig-inirad-temperature-duration} would be deeper while the peaks would be higher corresponding to the $ \theta $'s that can realize strong tachyonic instability in the first (or at least early) scalaron oscillation after the end of inflation, with smaller wiggles corresponding to late-time critical cases. Similar argument also applies to the right panel. 

\subsection{In the absence of tachyonic preheating}\label{Subsec-without-tachyonic}

In this subsection, we focus on the case where tachyonic instability is very weak or does not take place during preheating. As a result, there is practically no $ \rho_{\rm rad} $ initially. In this case, we can simply change the initial conditions for $ \rho_{\rm rad} $ and $ \rho_{\rm bg} $ in the numerical calculation in the previous subsection. 

\begin{figure}[h]
	\centering
	\begin{subfigure}
		\centering
		\includegraphics[width=.45\textwidth]{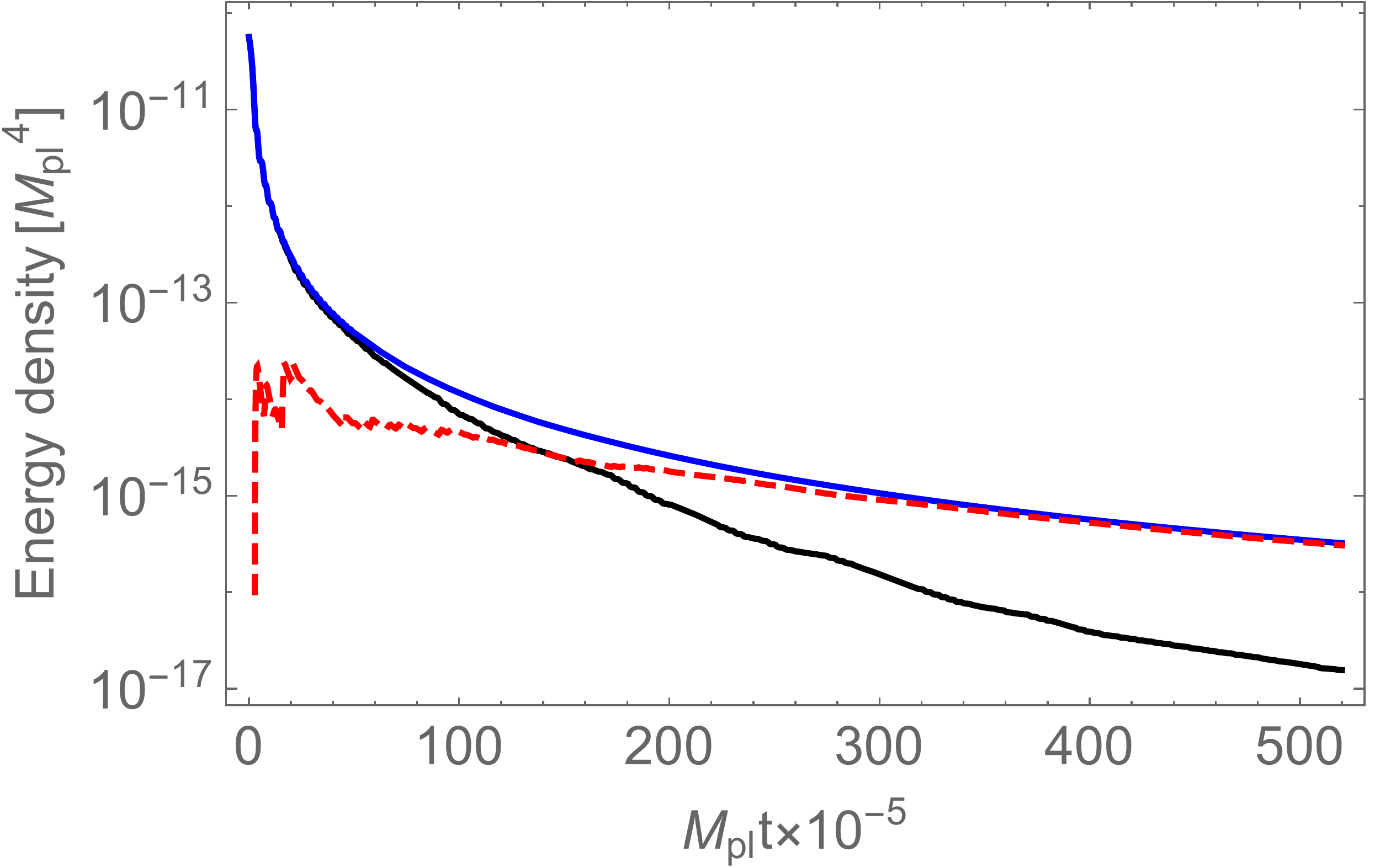}
	\end{subfigure}
	\begin{subfigure}
		\centering
		\includegraphics[width=.45\textwidth]{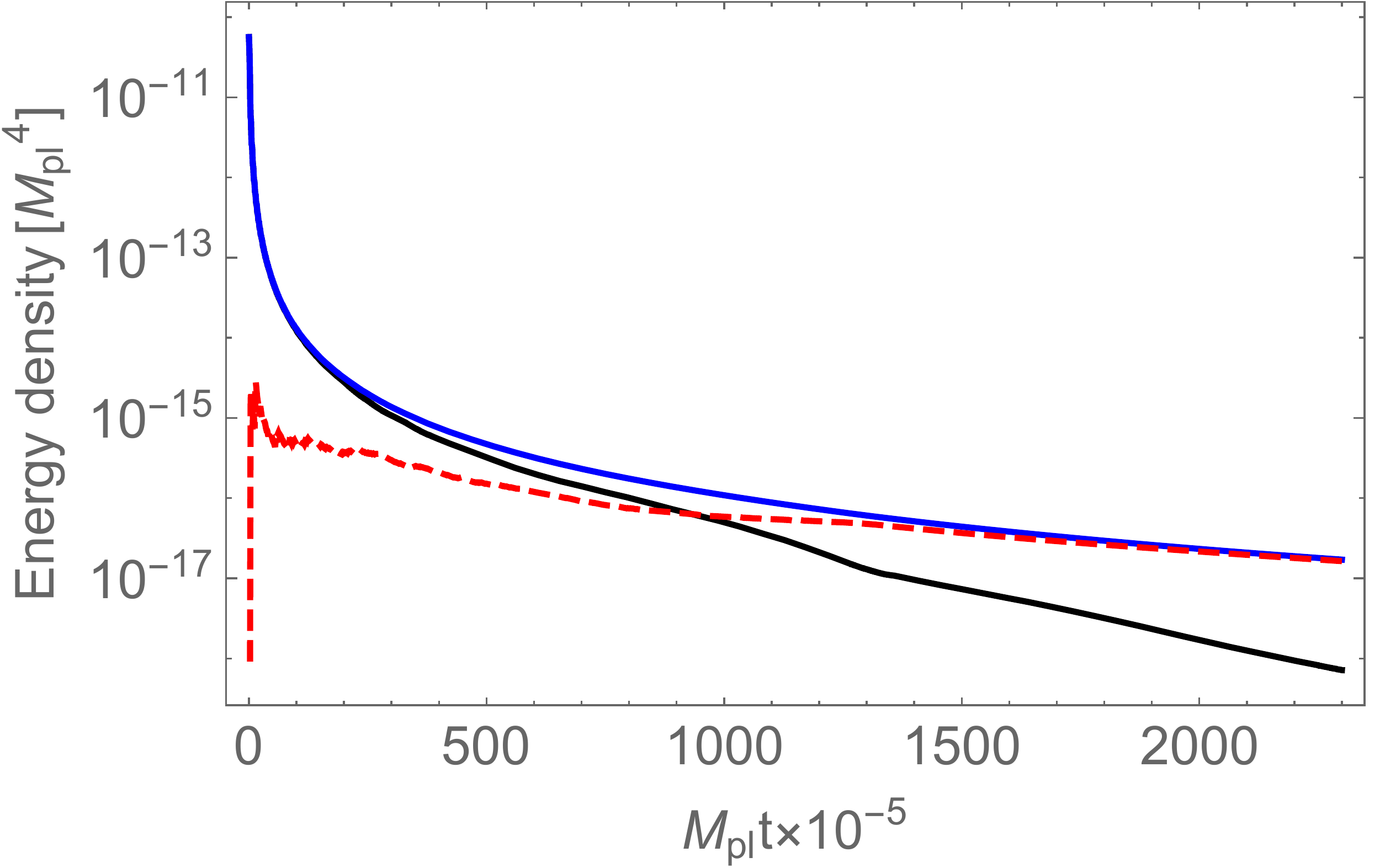}
	\end{subfigure}
	\begin{subfigure}
		\centering
		\includegraphics[width=.45\textwidth]{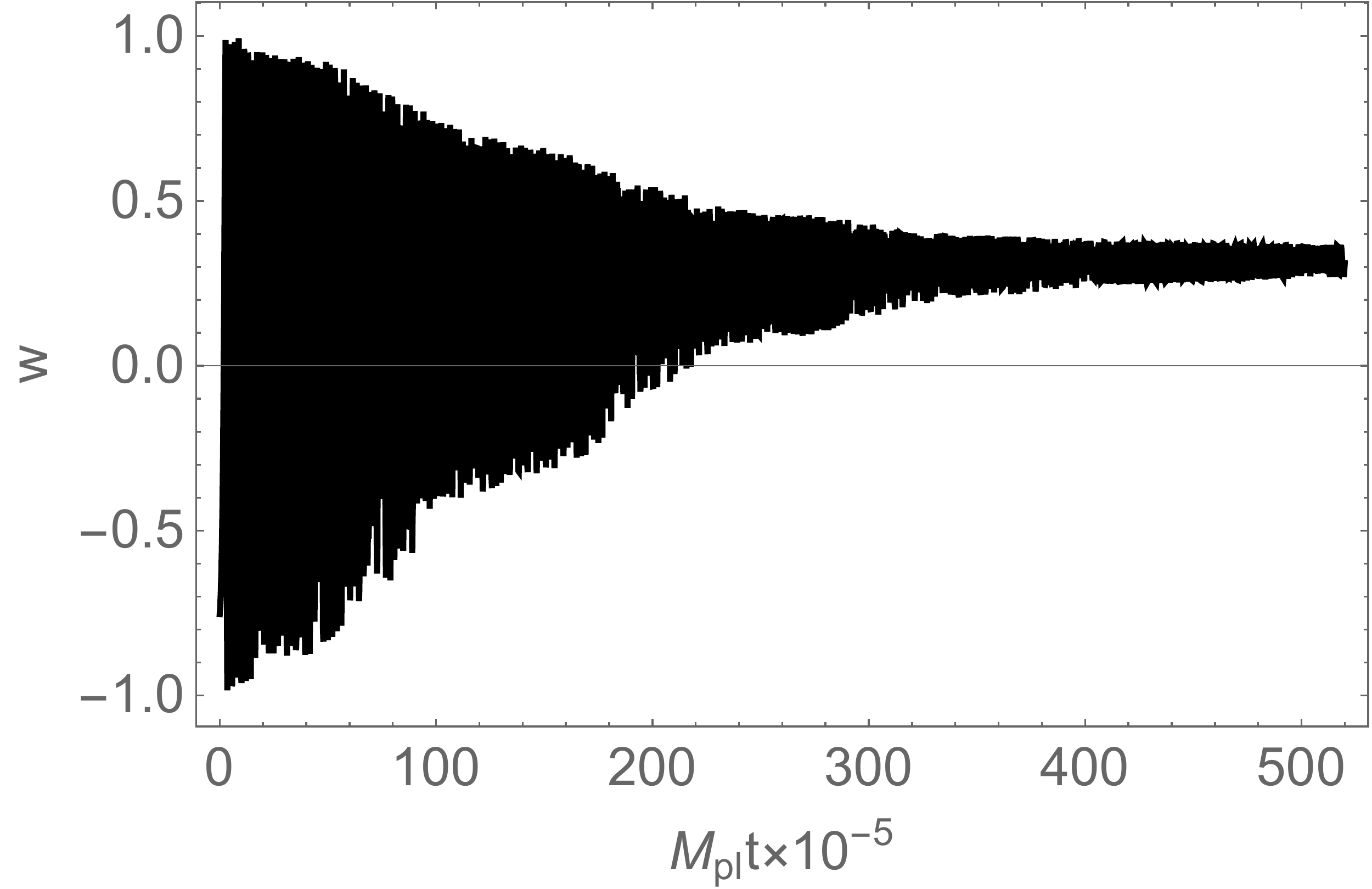}
	\end{subfigure}
	\begin{subfigure}
		\centering
		\includegraphics[width=.45\textwidth]{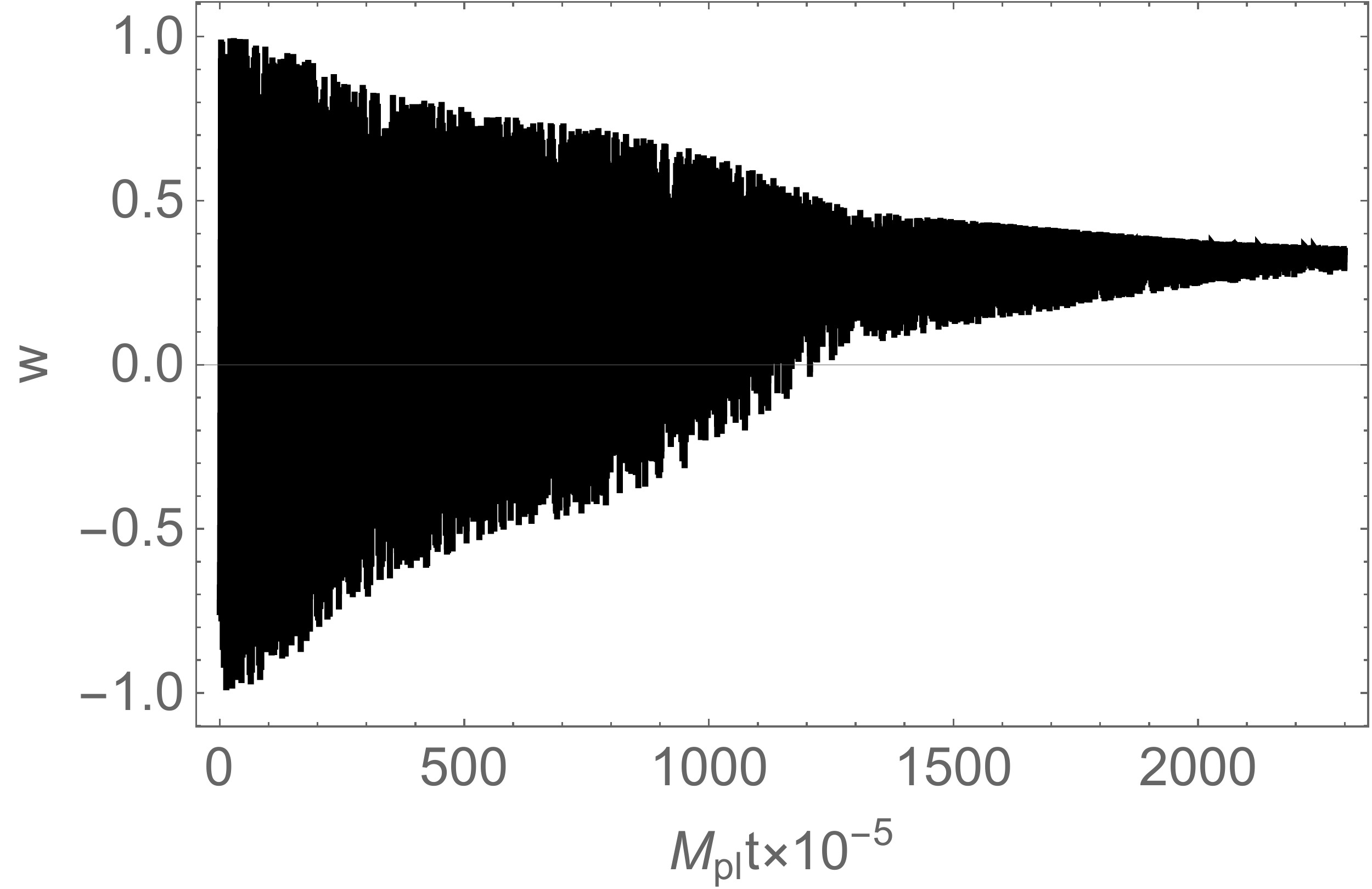}
	\end{subfigure}
	\caption{Up: Two examples of the evolution of $ \rho_{\rm rad} $ (red dashed), $ \rho_{\rm bg} $ (black solid), and $ 3M^2_{\rm pl} H^2 $ (blue solid) with $ \xi =3000 $ (left) and $ \xi=1000 $ (right) and initial radiation $ \rho_{\rm rad,ini} =0 $. Down: Two examples of the evolution of $ w $ up to $ \rho_{\rm rad}/\rho_{\rm bg} \gsim 20 $ with the same parameters as their upper panels respectively.}
	\label{fig-norad-energy-w}
\end{figure}

First, we show two examples of the evolution of energy densities and equation of state parameter in Fig.~\ref{fig-norad-energy-w}. 
As we can see there, even if there is no initial radiation before the onset of the perturbative reheating, the perturbative decay of scalaron and Higgs can still reheat the universe in a similar manner as the case in previous subsection. Still, we define the end of reheating as the moment when the condition $ \rho_{\rm rad}/\rho_{\rm bg} \gsim 20 $ becomes satisfied. 

\begin{figure}[h]
	\centering
	\begin{subfigure}
		\centering
		\includegraphics[width=.45\textwidth]{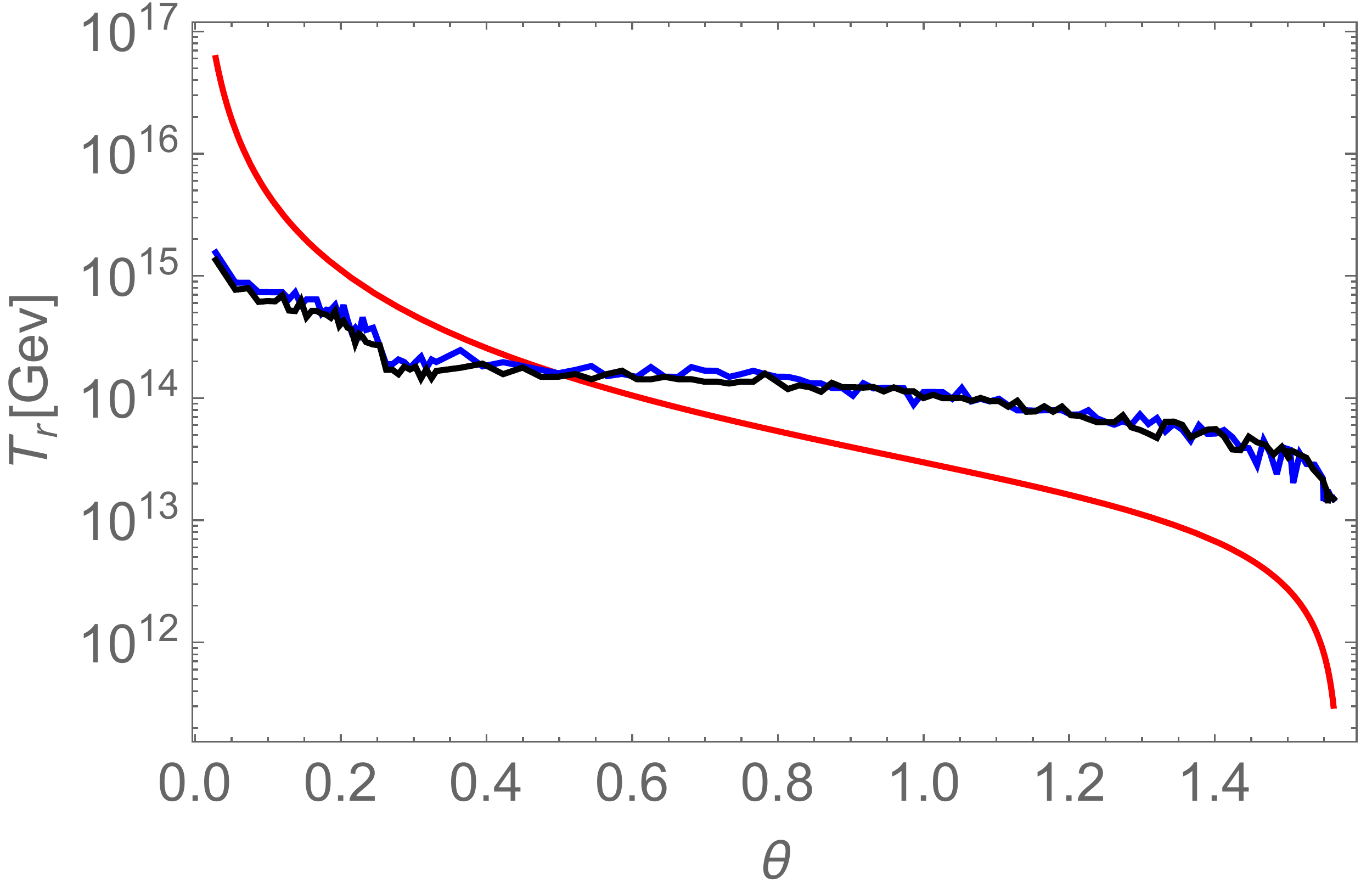}
	\end{subfigure}
	\begin{subfigure}
		\centering
		\includegraphics[width=.45\textwidth]{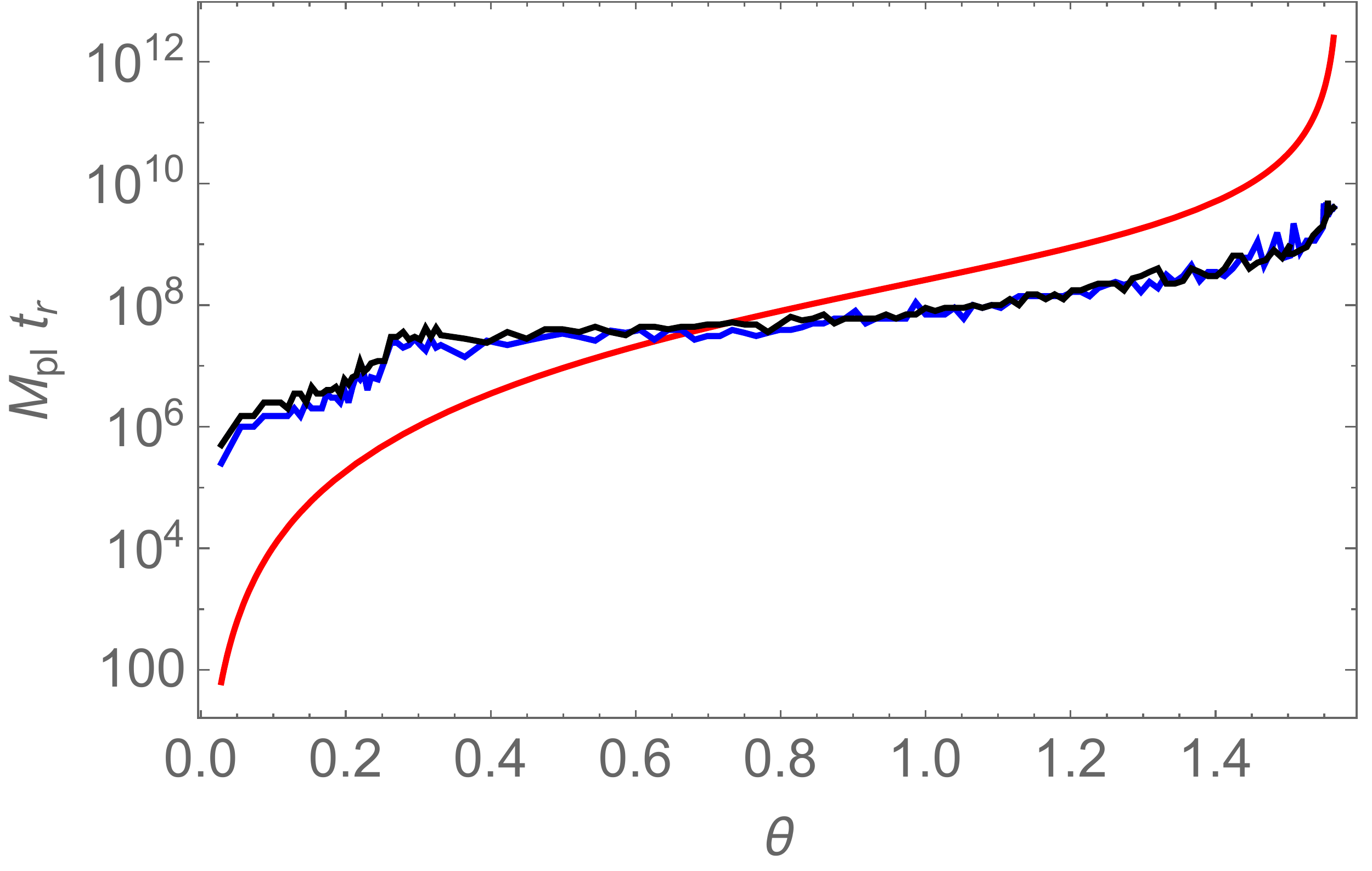}
	\end{subfigure}
	\caption{Left: reheating temperature $ T_r $ with (blue) and without (black) initial radiation, and the temperature calculated with only scalaron decay rate $ T_{\varphi} = 0.2\sqrt{\Gamma_{\varphi} M_{\rm pl}} $ (red). Right: duration of perturbative reheating $ t_r $ with (blue) and without (black) initial radiation, and the scalaron decay time $ \Gamma_{\varphi}^{-1} $ (red). Both $ T_r $ and $ t_r $ are determined at the moment when $ \rho_{\rm bg} $ becomes smaller than $ 5 \% $ of $ \rho_{\rm rad} $. The blue lines come from the results in Fig.~\ref{fig-inirad-temperature-duration} for convenient comparison between the cases with and without initial radiation energy.}
	\label{fig-norad-temperature-duration}
\end{figure}

Next, we discuss the reheating temperature $ T_r $ and duration $ t_r $ in the current setup. We again scan the parameter space $ 35 \leq \xi \leq \xi_s $ for the same reason discussed in Sec.~\ref{Subsec-with-tachyonic}. The results are shown in Fig.~\ref{fig-norad-temperature-duration} together with the results in Fig.~\ref{fig-inirad-temperature-duration} for convenient comparison. This is also one of the main results of this paper. 
As can be seen in Fig.~\ref{fig-norad-temperature-duration}, the results from cases with and without initial radiation energy almost coincide with each other, which indicates the fact that, if preheating processes cannot completely deplete the energy of the homogeneous background fields, once the homogeneous fields dominates the universe again and perturbative decay takes over the reheating process, the resulting reheating temperature (as well as the duration of reheating) is almost independent of the detail of preheating. Therefore, the perturbative decay indeed plays an important role in the whole reheating process in the mixed Higgs-$ R^2 $ model. 

\begin{figure}
	\centering
	\includegraphics[width=.45\textwidth]{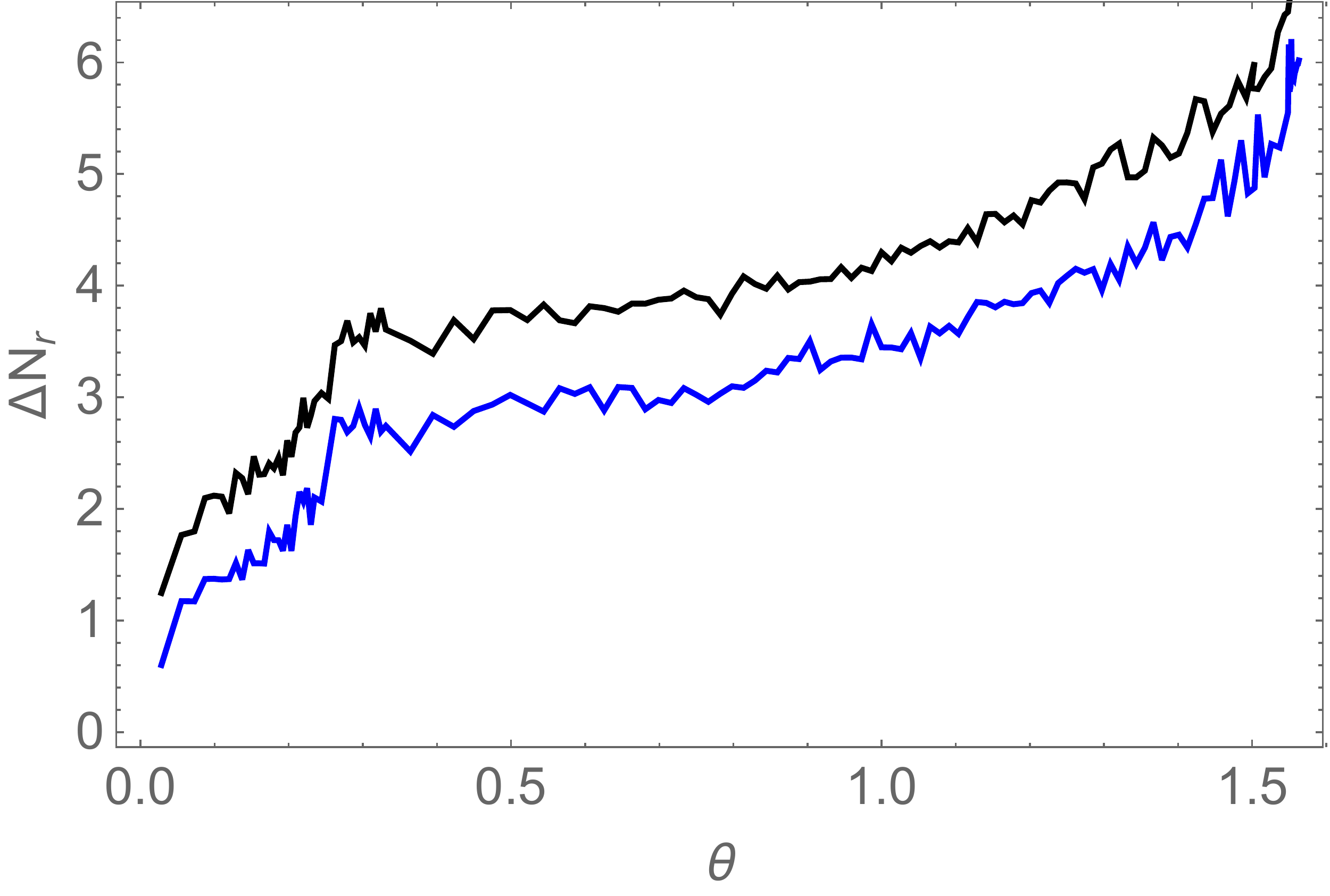}
	\caption{Comparison of e-fold numbers of cosmic expansion between the end of inflation and the end of reheating. Blue: number of e-fold in the case with initial radiation. Black: number of e-fold in the case without initial radiation.}
	\label{fig-compare-e-fold}
\end{figure}

Finally, we calculate the number of e-folds of reheating $ \Delta N_r $ which is important to confront theoretical inflation predictions with observational data. Figure~\ref{fig-compare-e-fold} compares $ \Delta N_r $ resulting from the cases with and without initial radiation. 
It can be seen that the difference is stable on a value slightly less than unity, which means that if there is a large fraction of energy in relativistic particles before the domination of perturbative reheating, the number of e-folds $ \Delta N_r $ needed to complete reheating can be reduced by around unity. Together with the duration of the early preheating stage itself, one may use the information of e-fold number to identify whether early preheating stage occurs or not. And the typical value of $ \Delta N_r $ shown in Fig.~\ref{fig-compare-e-fold} is $ 3 \lsim \Delta N_r \lsim 5 $. 

\section{Discussion and outlook}
\label{Sec-4}

In this paper, we have investigated the perturbative reheating in the mixed Higgs-$ R^2 $ inflation and discussed the role of perturbative decay throughout the whole reheating process. 

First of all, we would like to clarify our assumptions and caveats in this work. We mainly focus on parameters $ \xi_s \geq \xi \gg \mathcal{O}(10) $ for the following reasons. For $ \xi \lsim \mathcal{O}(10) $, the hierarchy between scalaron and Higgs oscillation becomes small so WKB approximation breaks down, which makes it difficult to study the system analytically. In addition, the perturbative reheating process takes long time to complete, which is challenging for numerical calculation. From the results presented above, one would expect that the system will quickly approach to the $ R^2 $-inflation limit as $ \xi $ decreases. Therefore, we do not discuss $ \xi \lsim \mathcal{O}(10) $ in detail in the present paper but only with some simple argument. Besides, we do not directly calculate tachyonic particle production but simply assume that the initial radiation energy in Sec.~\ref{Subsec-with-tachyonic} is produced by tachyonic instability during the early preheating stage. When defining the completion of reheating, we take the criterion $ \rho_{\rm rad}/\rho_{\rm bg} >20 $ instead of waiting for a long time $ \sim \mathcal{O}(\Gamma_{\varphi}^{-1}) $. We expect that our results are robust in the sense that the reheating temperature will not have order of magnitude difference when shifting the completion time later. As for the produced particles, we assume that they are relativistic and reach thermal equilibrium right at the end of reheating defined in this paper. Actually, when Higgs background decays in $ \varphi <0 $ regime, the produce fermions (top and bottom quarks in our consideration) are massless therefore relativistic. However, they gain masses as scalaron enters $ \varphi >0 $ regime and Higgs enters one of the valleys. Therefore, they are switching between being massless and massive during the oscillation of the homogeneous fields. When they become heavier during $ \varphi >0 $, they can further decay into other lighter particles. We do not study this phenomenon in detail in this work. 

Next, we summarize the conclusion of our paper. We first analytically investigated the evolution of the inflaton fields with WKB approximation, and discussed the relation between the energy distribution of the background fields and the parameter choice, which creates the fine structure of our results in Figs.~\ref{fig-inirad-temperature-duration},~\ref{fig-norad-temperature-duration}, and~\ref{fig-compare-e-fold}. Then we have calculated the perturbative decay rates for both homogeneous scalaron and Higgs in the WKB regime where the time scale of scalaron oscillation is much larger than that of Higgs, from which we estimated the possible effects of the decay on the period during and after the occurrence of tachyonic preheating. We have found that the decay of Higgs particles produced by tachyonic preheating may weaken the rescattering process between homogeneous and inhomogeneous modes in a non-negligible way, which could slow down the preheating process and allow the redomination of the oscillating background field. More importantly, we studied the perturbative reheating which determines the final reheating temperature $ T_r $ in this model. It is shown that larger $ \xi $ leads to larger $ T_r $ due to larger decay rates and more allowed decay channels. Also, the reheating temperature is typically larger than $ T_{\varphi} $ because the decay of homogeneous Higgs field increases the effective decay rate of the whole system. We take into account the early preheating stage by imposing non-vanishing radiation energy density to the initial conditions of the system. We find that early preheating stage can hardly affect the final reheating temperature but still can change the number of e-folds of the perturbative reheating at late time. 

This work only focuses on the case $ \xi \gg \mathcal{O}(10) $. For small $ \xi \lsim \mathcal{O}(10) $, we argue that the situation would quickly approach to $ R^2 $-limit, but more detailed investigation should be done in the future work in this parameter range. Also, this work only incorporated effects of tachyonic preheating phenomenologically in the sense that the effects of early preheating stage has been taken into account only as the extra radiation at the initial time of the numerical calculation. 
In order to trace the evolution of the system during preheating, lattice calculation is desirable, but is difficult to perform sufficiently long time and to extract the remaining amplitude of the homogeneous mode of the scalaron, which would dominate later to govern the final stage of reheating. 
We also leave it as a future work.

\section*{Acknowledgments}

The author is especially grateful to Professor Jun'ichi Yokoyama for his encouragement and great help. The author also thank Ryusuke Jinno, Kohei Kamada, Alexei A. Starobinsky for useful discussions, and Fedor Bezrukov, Chris Shepherd, Seong Chan Park and the referee for valuable comments. This work was supported by the Global Science Graduate Course (GSGC) program of the University of Tokyo and the JSPS Research Fellowships for Young Scientists.

\appendix 

\section{Energy distribution between Higgs and scalaron}\label{Appendix-kinetic-Higgs}

In this appendix, we discuss the energy distribution between Higgs and scalaron during (p)reheating. This is an significant point on at least two aspects considered here. Firstly, in Refs.~\cite{He:2018mgb,He:2020ivk}, we used a simple trick to calculate the amplitude of the first oscillation of the homogeneous background so that we can analytically calculate the particle production of spiky preheating and tachyonic preheating. This relies on the fact that the kinetic energy of Higgs is negligible when $ \varphi $ reaches its maximal amplitude during the oscillation $ \dot{\varphi}=0 $. It can be true because $ h\sim 0 $ in both cases. Secondly, as discussed in the main text of this paper, the energy distribution is important to determine the efficiency of the perturbative decay for different parameter choices. The reason is as follows. The decay rate of Higgs is much larger than that of scalaron. If a large fraction of energy is stored in the Higgs field, it is expected that the decay would be more efficient than the case where most energy is stored in the scalaron. As will be shown below, the closer to (but not coinciding with\footnote{If backreaction is taken into account, tachyonic instability in the critical case might be terminated midway, which could cause large amplitude of Higgs oscillation and so large energy stored in Higgs field. But we do not consider this case in the current paper.}) the critical parameters that can realize strong tachyonic preheating in the first scalaron oscillation after the end of inflation, the more energy is stored in the Higgs field, which results in a more efficient perturbative reheating and higher reheating temperature as discussed in the main text. Therefore, as we scan the parameter space to obtain the reheating temperature from perturbative decay, one can see dips and peaks due to the changing ``distance" from the critical parameters. 

We separate this discussion into two parts for (1) $ \varphi <0 $ regime in the first scalaron oscillation, and (2) subsequent oscillations, respectively, and restrict ourselves to $ \xi \gg \mathcal{O}(10) $ because it could be done analytically for the reason mentioned in Sec.~\ref{Subsec-bg}. Since Higgs is approximately a harmonic oscillator in this case, we can use its kinetic energy to characterize the total energy stored in Higgs. 

\subsection{$ \varphi <0 $ regime in the first scalaron oscillation}

In this case, the kinetic energy of Higgs is subdominant compared with the potential energy for most of time, which can be shown as follows. The effective mass of Higgs is given in Eq.~\eqref{eq-higgs-effmass} while the typical amplitude is $ h_{\rm osc} \sim M_{\rm pl}/\xi_c $. Therefore, the averaged kinetic energy can be estimated as 
\begin{align}
    E_{h,\rm kin} \sim m^2_{h_{\rm osc}} \frac{M^2_{\rm pl}}{\xi^2_c} \sim \frac{\xi}{\xi^2_c} M^2 M^2_{\rm pl} \alpha |\bar{\varphi}| ~.
\end{align}
where $ \bar{\varphi} $ denotes the time averaged value of $ \varphi $ during this period. On the other hand, the potential energy density is given by 
\begin{align}
    U (\varphi<0, h\simeq 0) \sim M^2_{\rm pl} M^2 \left( \alpha \bar{\varphi} \right)^2 ~.
\end{align}
The ratio between the kinetic and potential energy is then 
\begin{align}
\frac{E_{h,\rm kin}}{U (\varphi<0, h\simeq 0)} \sim \frac{\xi}{\xi^2_c} \frac{1}{\alpha |\bar{\varphi}|} \ll 1 
\end{align}
which is consistent with our expectation. Therefore, we can safely use the conclusion in Refs.~\cite{He:2018mgb,He:2020ivk} that the amplitude of $ \varphi $ when $ \varphi<0 $ for the first time is given by 
\begin{align}\label{eq-scalaron-varphi-1}
\alpha |\varphi_1| =C_1 \frac{M_c}{M} 
\end{align}
with $ C_1\simeq 0.25 $ taking into account the energy dissipation during inflation. 

\subsection{Subsequent oscillations}

The energy of homogeneous Higgs field in the subsequent oscillations are largely affected by the period $ \varphi >0 $ in the first scalaron oscillation. The argument is based on the special shape of the potential shown in Fig.~\ref{fig-potential}. If one starts with a nearly-critical parameter, by climbing up the potential hill around $ h=0 $ in $ \varphi >0 $ regime and falling down to one of the valleys during oscillation, scalaron can easily generate large amplitude of Higgs oscillation $ |h_{\rm osc}| \sim |h_{\rm v}| $ around the valley, which means that more energy is transferred to the Higgs. On the contrary, it is difficult for the Higgs to ``return" energy to the scalaron as long as the Higgs stays in the valleys when $ \varphi>0 $ in later evolution, which is the general case because fine-tuning is needed to climb up the hill (again)~\cite{He:2020ivk}. Consequently, if a large fraction of energy is transferred to the Higgs field in the first scalaron oscillation, this amount of energy will decay through Higgs, which is efficient. If one starts with a far-from-critical parameter, the oscillation amplitude of Higgs stays small long enough until a opportunity for the inflaton to climb up the hill in some subsequent oscillations. After that moment, the perturbative decay process can become more efficient, but the whole reheating period must be longer than the former case because the possible maximal efficiency of the tachyonic instability decreases as the amplitude of scalaron oscillation gets smaller. Thus, in the following, we focus on the regime $ \varphi >0 $ in the latter part of the first scalaron oscillation. 

Firstly, we consider the case where $ |h_{\rm osc}| \sim |h_{\rm v}| $ when the model parameter is nearly-critical. Specifically, we have the kinetic energy of Higgs 
\begin{align}
    E_{h, \rm kin} \sim m^2_{h_{\rm osc}} h^2_{\rm osc} \sim m^2_{h_{\rm osc}} h^2_{\rm v} \sim \mathcal{O}(10) \times \frac{\xi^2}{\lambda} \tilde{M}^2 M^2 \left( \alpha \varphi \right)^2
\end{align}
while the potential energy is given by 
\begin{align}
    U(\varphi>0, h\simeq h_{\rm v}) \sim M^2_{\rm pl} \tilde{M}^2 \left(\alpha\varphi \right)^2 ~.
\end{align}
As a result, the ratio between them is given by 
\begin{align}\label{eq-positive-varphi-kin-pot-ratio}
    \frac{E_{h,\rm kin}}{U (\varphi>0, h\simeq h_{\rm v})} \sim \mathcal{O}(10) \times \frac{\xi^2}{\lambda} \frac{M^2}{M^2_{\rm pl}} = \mathcal{O}(10) \times \frac{1}{3} \left( \frac{M^2}{\tilde{M}^2} -1 \right) ~.
\end{align}
For the case of interest $ \xi \gg \mathcal{O}(10) $, this could be much larger than order of unity, which means that a large fraction of energy is stored in the homogeneous Higgs field. Of course, the calculation above is just order estimate because generally speaking $ |h_{\rm osc}| $ cannot really exceed or equal $ |h_{\rm v}| $. Even if it does in some specific case, it is beyond the scope of this paper because we must talk about tachyonic preheating when $ |h_{\rm osc}| \gsim |h_{\rm v}|/2 $. 

Next, we focus on the case where $ |h_{\rm osc}| \ll |h_{\rm v}| $ when the model parameter is far from being critical. We repeat the calculation above while setting $ |h_{\rm osc}|\sim M_{\rm pl}/\xi_c $, 
\begin{align}
    \frac{E_{h,\rm kin}}{U(\varphi >0, h\simeq h_{\rm v})} \sim \frac{m^2_{h_{\rm osc}} M^2_{\rm pl}/\xi^2_c}{M^2_{\rm pl} \tilde{M}^2 (\alpha\varphi)^2} \sim \frac{\xi}{\xi^2_c} \frac{M^2}{\tilde{M}^2} \frac{1}{\alpha \varphi} \leq \left. \frac{\xi}{\xi^2_c} \frac{M^2}{\tilde{M}^2} \frac{1}{\alpha \varphi} \right|_{\xi =\xi_s} \simeq \dfrac{0.3}{\alpha \varphi} \sim \mathcal{O}(1)
\end{align}
which tells us that this ratio is at most of order unity. 

In conclusion, there is a larger fraction of energy stored in Higgs for model parameters closer to critical cases while a smaller fraction for parameters far from the critical points. 

\section{Analytical calculation of the evolution of homogeneous Higgs}\label{Appendix-analytical-bg}

In this appendix, we will derive the analytical expression for the Higgs background field in the case $ \xi \gg \mathcal{O}(10) $ with WKB approximation and the information in Sec.~\ref{Subsubsec-bg-Higgs}. 

\subsection{$ \xi \gg \mathcal{O}(10) $}

\begin{figure}[h]
	\centering
	\begin{subfigure}
		\centering
		\includegraphics[width=.45\textwidth]{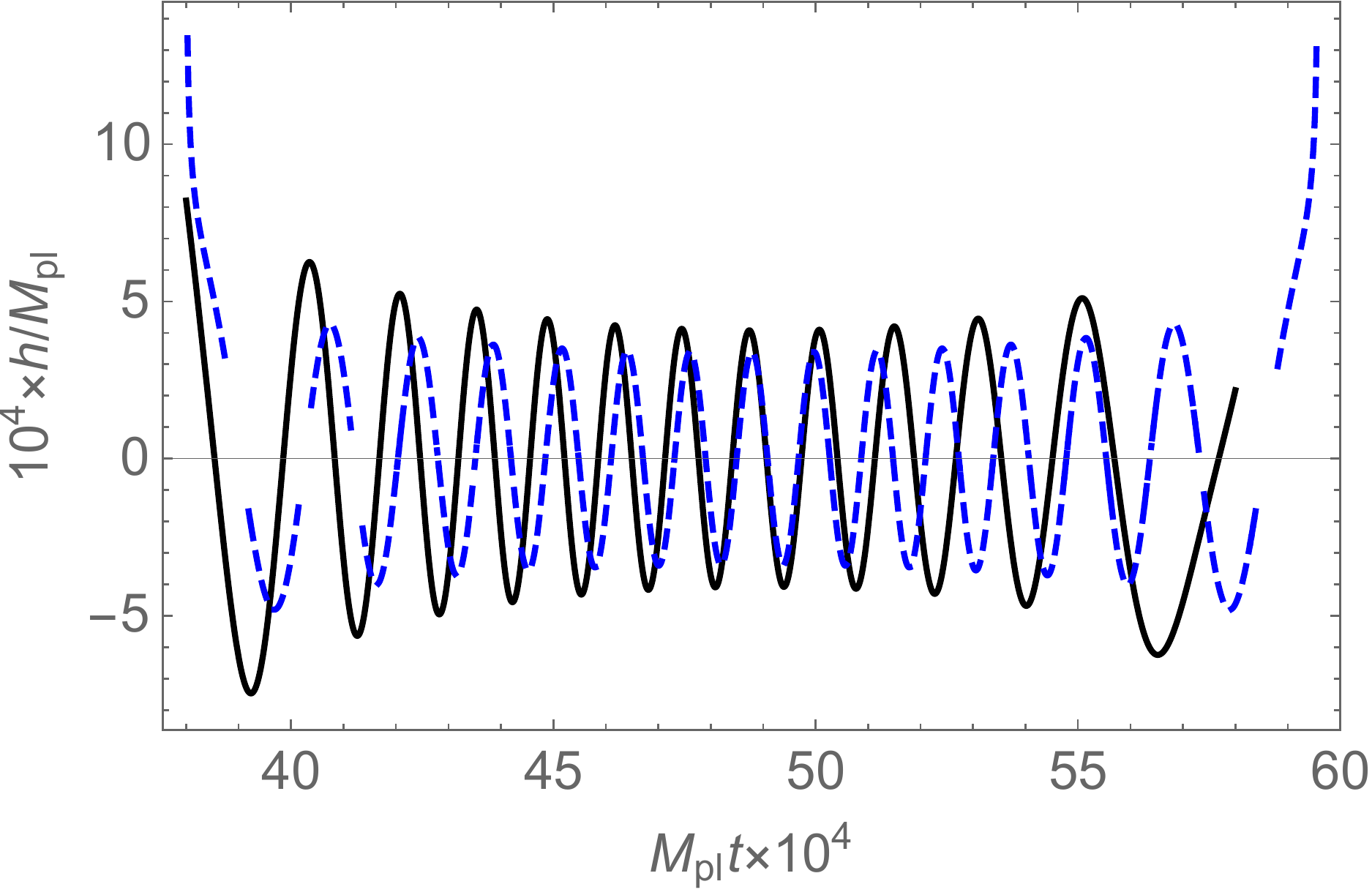}
	\end{subfigure}
	\begin{subfigure}
		\centering
		\includegraphics[width=.45\textwidth]{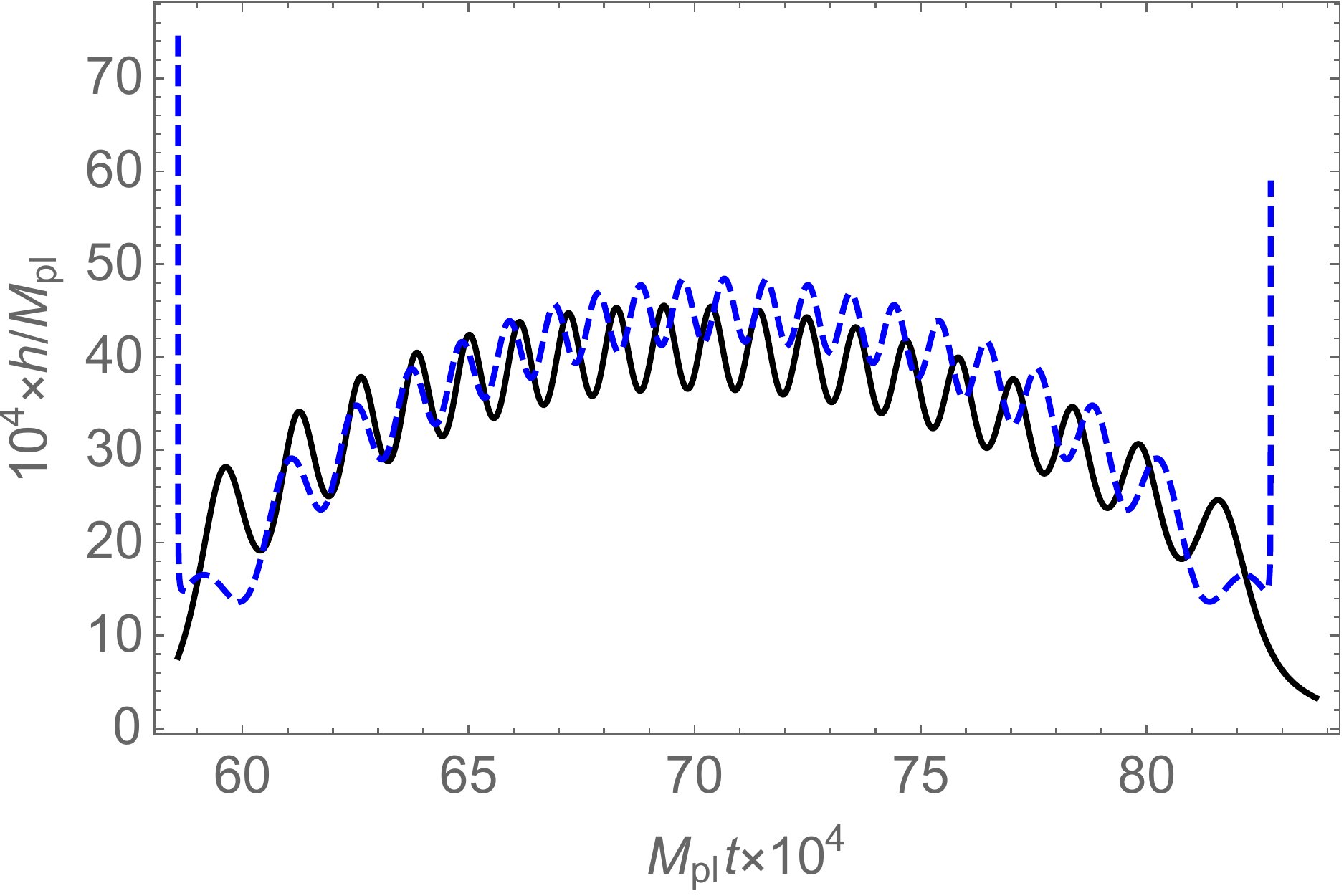}
	\end{subfigure}
	\caption{Comparison of the analytical and numerical solution of the background evolution of Higgs with $ \xi=2000 $ as an example. Left: during $ \varphi<0 $. Right: during $ \varphi >0 $. Solid: numerical solution of the exact equation of motion of Higgs background. Dashed: analytical solution with WKB approximation. Here we fix $ A_h$ so that the amplitude of $ h_{\rm osc} \simeq 1.5 \times M_{\rm pl}/\xi_c $.}
	\label{fig-bg-higgs-analytic}
\end{figure}

In this case, WKB approximation is valid because $ m^2_{h_{\rm osc}} \gg m^2_{\varphi} $ is satisfied. We write down Eq.~\eqref{eq-bg-higgs-WKB} here again for convenience
\begin{align}\label{eq-bg-higgs-WKB-app}
h_{\rm osc} = \frac{A_h}{\sqrt{2m_{h_{\rm osc}}}}  \cos\left[\int^t m_{h_{\rm osc}} (t') dt' +\delta_i \right] 
\end{align}
where $ \delta_i $ is the phase determined by the initial condition and $ A_h $ is a constant determining the amplitude. During one oscillation of $ \varphi $, $ h_{\rm osc} $ oscillates many times so we can further neglect the cosmic expansion. Therefore, from Sec.~\ref{Subsubsec-bg-scalaron} we know that $ \varphi \propto \sin (Mt) $ or $ \varphi \propto \sin (\tilde{M}t) $ for $ \varphi <0 $ and $ \varphi >0 $ regimes respectively. As a result, the integration can be done analytically 
\begin{align}
    \int^t m_{h_{\rm osc}}(t') dt' = \left\{
    \begin{array}{ll}
        -2 \sqrt{3\xi \alpha |\varphi_a|} E\left[ \frac{\pi}{4} -\frac{Mt}{2}, 2 \right] &, \varphi <0 \\ [0.3cm]
        -2 \sqrt{6\xi \alpha \varphi_a} \frac{M}{\tilde{M}} E\left[ \frac{\pi}{4} -\frac{\tilde{M}t}{2}, 2 \right] &, \varphi >0
    \end{array}
    \right.
\end{align}
where $ E[y,x] =\int^{y}_0 \left( 1-x \sin^2 t \right)^{1/2} dt $ is the elliptic integral of the second kind and $ \varphi_a $ is defined as $ \varphi =\varphi_a \sin \left( M(\varphi) t \right) $ for both positive and negative $ \varphi $. Putting this back into Eq.~\eqref{eq-bg-higgs-WKB-app} gives the analytical solution of $ h_{\rm osc} $ which is not a bad approximation although it overestimates the frequency slightly (see Fig.~\ref{fig-bg-higgs-analytic} for example). The error could come from the estimation of the amplitude of $ \varphi $ in Eq.~\eqref{eq-scalaron-varphi-1}. 

\subsection{$ \xi \lsim \mathcal{O}(10) $}

The condition $ \xi \lsim \mathcal{O}(10) $ means that the model is very close to the $ R^2 $-limit. In this case, the valleys in $ \varphi >0 $ regime become shallow and approach to merge, as can be seen from Eq.~\eqref{eq-h-valley} that $ h_{\rm v} $ is proportional to $ \xi $ when $ \varphi >0 $. Besides, the effective mass of $ h_{\rm osc} $ becomes comparable with that of $ \varphi $ so that the huge hierarchy between them presented in previous subsection disappears. As a result, the perturbative decay process of the Higgs field becomes much less efficient so the reheating efficiency of the whole system becomes close to $ R^2 $-inflation. It is difficult to derive the analytical solution in this case and we do not intend to do that in this paper. 

\section{Derivation of the decay rates for Higgs and scalaron}\label{Appendix-decay-rate}

In this appendix, we will present the derivation of the decay rates for both Higgs and scalaron fields in the WKB regime when the scalaron motion is sufficiently slow in the time scale of the Higgs oscillation. We only consider the decay of Higgs into $ W^{\pm} $, $ Z $ bosons and top, bottom quarks, as well as the decay of scalaron into Higgs particles. We work in unitary gauge where $ \mathcal{H} =h/\sqrt{2} $. 

\subsection{Higgs decay rate}\label{Subsec-Higgs-decay-rate} 

In the Jordan frame, the relevant sector is written as 
\begin{align}\label{eq-jordan-higgs-gauge-fermion}
    S_{\mathrm J} \supset \int\!\! d^4 x \sqrt{-g_{\mathrm J}} \left[ \xi |\mathcal{H}|^2 R_{\rm J} - g_{\mathrm J}^{\mu\nu} \left( D_{{\rm J}\mu}  {\mathcal H} \right)^\dagger D_{{\rm J}\nu} {\mathcal H} -\lambda |\mathcal{H}|^4 + K_{\rm J, gauge} -\frac{y_t}{\sqrt{2}} h \bar{t}t - \frac{y_b}{\sqrt{2}} h \bar{b}b  \right]
\end{align}
where $ D_{{\rm J} \mu} =\nabla _{{\rm J} \mu} +\frac{i}{2} g' B_{\mu} +i g W^a_{\mu} \tau^a $ is the covariant derivative with $ \sigma^a =2\tau^a $ the Pauli matrices, and $ K_{\rm J, gauge} $ is the gauge invariant kinetic terms of the gauge fields given as 
\begin{align}
    K_{\rm J, gauge} =&-\frac{1}{4} g^{\mu \rho}_{\rm J} g^{\nu \sigma}_{\rm J} B_{\mu \nu} B_{\rho \sigma} -\frac{1}{4} g^{\mu \rho}_{\rm J} g^{\nu \sigma}_{\rm J} W^a_{\mu \nu} W^a_{\rho \sigma} \label{eq-jordan-gauge-kinetic} \\
    B_{\mu\nu} \equiv & \partial_{\mu} B_{\nu} -\partial_{\nu} B_{\mu} \\
    W^a_{\mu\nu} \equiv & \partial_{\mu} W^a_{\nu} -\partial_{\nu} W^a_{\mu} -g \epsilon^{abc} W^b_{\mu} W^c_{\nu} ~,
\end{align}
and the last two terms for the mass terms for top and bottom quarks. By the one-loop renormalization group running (see for example Ref.~\cite{Buttazzo:2013uya}), we have the gauge couplings $ g \approx 0.55 $, $ g'\approx 0.42 $, and the Yukawa couplings $ y_t \approx 0.5 $, $ y_b \approx 0.01 $ at energy scale $ \sim 10^{12} $~GeV which corresponds to the energy scale at the end of inflation in this model. By the conformal transformation \eqref{eq-conformal-transformation}, we go to the Einstein frame 
\begin{align}\label{eq-einstein-higgs-gauge-fermion}
    S_{\mathrm{E}} \supset \int d^4 x \sqrt{-g} \left[ - e^{-\alpha \varphi} g^{\mu\nu} \left( D_{\mu}  {\mathcal H} \right)^\dagger D_{\nu} {\mathcal H} -U(\varphi, \mathcal{H}) +K_{\rm E, gauge} -e^{-2\alpha \varphi} \left( \frac{y_t}{\sqrt{2}} h \bar{t}t + \frac{y_b}{\sqrt{2}} h \bar{b}b \right) \right]
\end{align}
where $ D_{\mu} =\nabla _{\mu} +\frac{i}{2} g' B_{\mu} +i g W^a_{\mu} \tau^a $ is the covariant derivative, $ U(\varphi, \mathcal{H}) $ is given in Eq.~\eqref{eq-einstein-potential}, and $ K_{\rm E, gauge} $ is the kinetic terms of the gauge field in Einstein frame 
\begin{align}\label{eq-einstein-gauge-kinetic}
    K_{\rm E, gauge} =-\frac{1}{4} g^{\mu \rho} g^{\nu \sigma} B_{\mu \nu} B_{\rho \sigma} -\frac{1}{4} g^{\mu \rho} g^{\nu \sigma} W^a_{\mu \nu} W^a_{\rho \sigma}
\end{align}
which is apparently conformally invariant. By rotating the gauge fields, we define $ Z_{\mu} $, $ W^{\pm}_{\mu} $, and the photon $ A_{\mu} $ as 
\begin{align}\label{eq-define-AWZ}
    Z_{\mu} \equiv& \cos \theta_w W^3_{\mu} -\sin \theta_w B_{\mu} \\
    A_{\mu} \equiv & \sin \theta_w W^3_{\mu} +\cos \theta_w B_{\mu} \\
    W^{\pm}_{\mu} \equiv & \frac{1}{\sqrt{2}} \left( W^1_{\mu} \mp i W^2_{\mu} \right)
\end{align}
with $ \tan \theta_w =g'/g $. Keeping terms up to quadratic order and neglecting  the conformal factor $ \exp (-\alpha \varphi) $, one obtains the relevant terms for $ W^{\pm}_{\mu} $ and $ Z_{\mu} $ up to quadratic order in the unitary gauge $ \mathcal{H} =h/\sqrt{2} $
as 
\begin{align}
    \mathcal{L} \supset & - \frac{g^2}{8} g^{\mu\nu} h_0^2 \left( 2W^+_{\mu} W^-_{\nu} + \frac{Z_{\mu}Z_{\nu}}{\cos^2\theta_w} \right) -\frac{3\lambda}{2} \frac{M^2}{\tilde{M}^2} h_0^2 \delta h^2 + \frac{\xi M^2}{\alpha M^2_{\rm pl}} \varphi \delta h^2 
\end{align}
where we have treated $ h $ as the sum of homogeneous and inhomogeneous parts $ h=h_0 (t) +\delta h (t, x) $. 
We know that $ h_0(t) $ is given by Eq.~\eqref{eq-higgs-valley-plus-oscillation}. When $ \varphi >0 $, the approximation $ |h_{\rm v}| \gg |h_{\rm osc}| $ is valid so we obtain 
\begin{align}
    \mathcal{L} \supset & 
    - \frac{g^2}{8} g^{\mu\nu} h_{\rm v}^2 \left( 2W^+_{\mu} W^-_{\nu} + \frac{Z_{\mu}Z_{\nu}}{\cos^2\theta_w} \right) -\frac{3\lambda}{2} \frac{M^2}{\tilde{M}^2} h_{\rm v}^2 \delta h^2 + \frac{\xi M^2}{\alpha M^2_{\rm pl}} \varphi \delta h^2 \nonumber \\
    &- \frac{g^2}{4} g^{\mu\nu} h_{\rm v} h_{\rm osc} \left( 2W^+_{\mu} W^-_{\nu} + \frac{Z_{\mu}Z_{\nu}}{\cos^2\theta_w} \right) -3\lambda \frac{M^2}{\tilde{M}^2} h_{\rm v} h_{\rm osc} \delta h^2 ~. \label{eq-higgs-decay-gauge} 
\end{align}
When $ \varphi <0 $, $ h_{\rm v} =0 $ so we have 
\begin{align}
    \mathcal{L} \supset - \frac{g^2}{8} g^{\mu\nu} h_{\rm osc}^2 \left( 2W^+_{\mu} W^-_{\nu} + \frac{Z_{\mu}Z_{\nu}}{\cos^2\theta_w} \right) -\frac{3\lambda}{2} \frac{M^2}{\tilde{M}^2} h_{\rm osc}^2 \delta h^2 + \frac{\xi M^2}{\alpha M^2_{\rm pl}} \varphi \delta h^2 \label{eq-higgs-decay-gauge-no}
\end{align}
where there is no three-leg interaction terms for the homogeneous Higgs to decay so the homogeneous Higgs field can only decay when $ \varphi >0 $. Based on Eq.~\eqref{eq-higgs-decay-gauge}, we calculate the decay rates of the homogeneous Higgs field into $ W^{\pm} $ and $ Z $ bosons. Note that we do not take into account the non-Abelian nature of the gauge fields which is not important to the lowest order of gauge coupling. 

As seen from above, the first line of Eq.~\eqref{eq-higgs-decay-gauge} gives the effective masses for the gauge bosons and the Higgs particles. In FLRW background, $ \sqrt{-g} =a^3 $. We redefine the fields as $ \tilde{W}^{\pm}_{\mu} \equiv a^{3/2} W^{\pm}_{\mu} $, $ \tilde{Z}_{\mu} \equiv a^{3/2} Z_{\mu} $, and $ \tilde{\delta h}\equiv a^{3/2} \delta h $ so that we can write down their effective masses without the scale factor 
\begin{align}
    m^2_{\tilde{W}} &\equiv \frac{g^2}{4} h^2_{\rm v} \\
    m^2_{\tilde{Z}} &\equiv \frac{g^2}{4\cos^2\theta_w} h^2_{\rm v} =\frac{g^2+g'^2}{4} h^2_{\rm v} \\
    m^2_{\tilde{\delta h}} &\equiv 3\lambda \frac{M^2}{\tilde{M}^2} h^2_{\rm v} - 2\frac{\xi M^2}{\alpha M^2_{\rm pl}}\varphi 
\end{align}
with $ h_{\rm v} $ given in Eq.~\eqref{eq-h-valley}. Now we are ready for the calculation of the corresponding decay rates. Specifically, for $ \varphi >0 $, the decay rates are calculated as 
\begin{align}
    \Gamma_{h_0 \rightarrow \tilde{W}\tilde{W}} &= \frac{\lambda}{8\pi} \frac{M^3}{\tilde{M}^2} \left( 6\xi \alpha\varphi \right)^{1/2} \left( 1-\frac{g^2}{2\lambda} \frac{\tilde{M}^2}{M^2} + \frac{3g^4}{16\lambda^2} \frac{\tilde{M}^4}{M^4} \right) \left( 1-\frac{g^2}{2\lambda} \frac{\tilde{M}^2}{M^2} \right)^{1/2}  \label{eq-higgs-to-W} \\
    \Gamma_{h_0 \rightarrow \tilde{Z}\tilde{Z}} &= \frac{\lambda}{16\pi} \frac{M^3}{\tilde{M}^2} \left( 6\xi \alpha\varphi \right)^{1/2} \left( 1- \frac{g^2+g'^2}{2\lambda} \frac{\tilde{M}^2}{M^2} + \frac{3\left( g^2+g'^2 \right)^2}{16\lambda^2} \frac{\tilde{M}^4}{M^4} \right) \left( 1-\frac{g^2+g'^2}{2\lambda} \frac{\tilde{M}^2}{M^2} \right)^{1/2} ~. \label{eq-higgs-to-Z} 
\end{align}
These decay channels are open only when the conditions 
\begin{align}
    \xi^2 >& \xi_c^2 \left( 1-\frac{2\lambda}{g^2} \right) \equiv \xi_W^2 \simeq 4291^2 \\
    \xi^2 >& \xi_c^2 \left( 1-\frac{2\lambda}{g^2+g'^2} \right) \equiv \xi_Z^2 \simeq 4347^2 
\end{align}
are satisfied respectively, which means that the decay of homogeneous Higgs into $ W^{\pm} $ and $ Z $ is possible only in the deep Higgs-like regime. 

As for the fermion sector, the relevant interaction terms are 
\begin{align}\label{eq-higgs-quarks-interaction}
    {\cal L}_{\rm Yukawa} \supset& -\frac{y_t}{\sqrt{2}} h_{\rm v} \bar{t}t - \frac{y_b}{\sqrt{2}} h_{\rm v} \bar{b}b \nonumber \\ &-\frac{y_t}{\sqrt{2}} h_{\rm osc} \bar{t}t - \frac{y_b}{\sqrt{2}} h_{\rm osc} \bar{b}b \nonumber \\
    &-\frac{y_t}{\sqrt{2}} \delta h \bar{t}t - \frac{y_b}{\sqrt{2}} \delta h \bar{b}b
\end{align}
where we also neglect the $ \exp (-2\alpha\varphi) $ factor for leading order analysis. Redefining the quarks $ \tilde{t} \equiv a^{3/2} t $ and $ \tilde{b} \equiv a^{3/2} b $ to absorb the factor $ \sqrt{-g} $, the masses of the new fermion fields are given by the first line in Eq.~\eqref{eq-higgs-quarks-interaction} as 
\begin{align}
    m_{\tilde{t}}\equiv& \frac{y_t}{\sqrt{2}} h_{\rm v} \\
    m_{\tilde{b}}\equiv& \frac{y_b}{\sqrt{2}} h_{\rm v} ~.
\end{align}
Unlike the gauge sector, the Higgs background can decay into top and bottom quarks in both positive and negative $ \varphi $ regimes. The decay width can be calculated as 
\begin{align}
    \Gamma_{h_0 \rightarrow \tilde{t} \bar{\tilde{t}}} =&\left\{
    \begin{array}{ll} 
        \frac{3y^2_t}{16\pi} \left( 3\xi\alpha |\varphi| \right)^{1/2} M &, \varphi <0 \\ [0.3cm]
        \frac{3y^2_t}{16\pi} \left( 6\xi\alpha \varphi \right)^{1/2} M \left( 1- \frac{y^2_t}{\lambda} \frac{\tilde{M}^2}{M^2} \right)^{3/2} &, \varphi >0
    \end{array}
    \right. \\
    \Gamma_{h_0 \rightarrow \tilde{b} \bar{\tilde{b}}} =&\left\{
    \begin{array}{ll} 
        \frac{3y^2_b}{16\pi} \left( 3\xi\alpha |\varphi| \right)^{1/2} M &, \varphi <0 \\ [0.3cm]
        \frac{3y^2_b}{16\pi} \left( 6\xi\alpha \varphi \right)^{1/2} M \left( 1- \frac{y^2_b}{\lambda} \frac{\tilde{M}^2}{M^2} \right)^{3/2} &, \varphi >0
    \end{array}
    \right. ~.
\end{align}
Because the quarks are massless during $ \varphi <0 $, the channels are always open during that period. On the contrary, 
the decay to top quarks is allowed during $ \varphi >0 $ only when 
\begin{align}
    \xi^2 >& \xi^2_c \left( 1-\frac{\lambda}{y^2_t} \right) \equiv \xi_t^2 \simeq 4351^2 
\end{align}
while the bottom quark channel is always open due to the small coupling $ y_b $.

\subsection{Scalaron decay rate}\label{Subsec-scalaron-decay-rate} 

The dominant decay channel of scalaron for $ \xi \gg \mathcal{O}(10) $ is decaying into Higgs particles with which the scalaron directly couples. 
One can extract the relevant terms from Eq.~\eqref{eq-higgs-decay-gauge} which we write down here again for convenience 
\begin{align}
    \mathcal{L} \supset& ~ -\frac{3\lambda}{2} \frac{M^2}{\tilde{M}^2} h^2_{\rm v} \delta h^2 + \frac{\xi M^2}{\alpha M^2_{\rm pl}} \varphi \delta h^2 ~.
\end{align}
Scalaron can decay into $ \delta h $ through these terms and the decay rate can be computed as \cite{Starobinsky:1980te,Zeldovich:1977vgo,Starobinsky:1981vz,Vilenkin:1985md} 
\begin{align}
    \Gamma_{\varphi \rightarrow \tilde{\delta h} \tilde{\delta h}} 
    &=\left\{ 
    \begin{array}{ll}
        \frac{3}{16\pi} \frac{M^3}{M^2_{\rm pl}} \left( \frac{1}{6} +\xi \right)^2 \left( 1- 12\xi \alpha |\varphi| \right)^{1/2} &, \varphi <0 \\ [0.3cm] 
        \frac{3}{16\pi} \frac{\tilde{M}^3}{M^2_{\rm pl}} \left( \frac{1}{6}-2\xi \frac{M^2}{\tilde{M}^2} \right)^2 \left( 1- 24 \xi \frac{M^2}{\tilde{M}^2} \alpha \varphi \right)^{1/2} &, \varphi >0 ~. \label{eq-scalaron-to-higgs-2}
    \end{array}
    \right.
\end{align}
The decay during $ \varphi <0 $ vanishes when $ \xi =-1/6 $ as expected, which corresponds to the usual conformally coupled case. However, for $ \varphi >0 $ this result does not admit negative non-minimal coupling because we have assumed the existence of the valleys with finite $ |h_{\rm v}| $ which is inconsistent with $ \xi <0 $. Instead, the decay rate for $ \varphi >0 $ vanishes when $ \xi \simeq 1/12 $. The minimally coupled case $ \xi =0 $ does not mean the decay channel is closed. In this case, our result coincides with e.g. Refs.~\cite{Gorbunov:2010bn,Gorbunov:2012ns}. In this paper, we can omit the $ 1/6 $ because we focus on $ \xi \gg \mathcal{O}(10) $. Additionally, only when $ m_{\varphi} >2m_{\delta h} $, the decay is kinematically allowed. In the case where $ \xi \gg 1 $, the decay process can only occur when $ |\varphi| $ becomes sufficiently small, which can be realized in two possible situations. The first one is when $ \varphi $ gets close to the origin during oscillation. The second is at late time when $ |\varphi| $ becomes small due to the Hubble expansion and decay of energy through Higgs.

\bibliography{ref}{}

\end{document}